\documentclass[12pt]{article}
\usepackage{amsmath}
\usepackage{graphicx}
\usepackage{enumerate}
\usepackage{natbib}
\usepackage{url} 
\usepackage{amssymb}
\usepackage{amsthm}
\DeclareMathOperator*{\argmin}{arg\, min}
\usepackage{wrapfig}
\usepackage{mathtools}
\usepackage{float}
\usepackage[hidelinks]{hyperref}
\usepackage{mathtools}
\usepackage{caption}
\usepackage{subcaption}
\usepackage{scalerel,stackengine}
\usepackage{relsize}
\usepackage{mathrsfs}
\usepackage{algorithm}
\usepackage{algpseudocode}
\usepackage{xcolor}
\usepackage{changebar}
\newtheorem{theorem}{Theorem}

\newtheorem{proposition}{Proposition}
\newtheorem{remark}{Remark}
\newtheorem{definition}{Definition}

\newcommand\normx[1]{\left\Vert#1\right\Vert}
\newcommand\bmat[1]{\mathbf{#1}}
\newcommand\myprod[2]{#1^{T} #2}
\newcommand\tildesym[1]{\boldsymbol{\tilde{#1}}}
\newcommand{\nacf}[0]{\operatorname{nacf}}

\newcommand{\pnacf}[0]{\operatorname{pnacf}}
\newcommand{\GNAR}[0]{\operatorname{GNAR}}

\newcommand{\rstage}[0]{\mbox{$r$-stage}}
\newcommand{\hlag}[0]{\mbox{$h$-lag}}

\newcommand{\rscc}[0]{\operatorname{rscc}}
\newcommand{\local}[0]{\operatorname{local}}
\newcommand{\globindex}[0]{\operatorname{globindex}}

\newcommand{\blind}{1}

\begin{document}
\nocite{*}

\def\spacingset#1{\renewcommand{\baselinestretch}%
{#1}\small\normalsize} \spacingset{1}


\if1\blind
{
  \title{\bf New tools for network time series with an application
    to COVID-19  hospitalisations}
  \author{G. P. Nason\footnotemark[1] \footnotemark[2],
    D. Salnikov\thanks{Dept.\ Mathematics, Huxley Building, Imperial College, 180 Queen's Gate, South Kensington, London, SW7 2AZ, UK.} \footnotemark[2]\\
    Imperial College London\\
    and \\
    M. Cortina-Borja\thanks{Great Ormond Street Institute of Child Health, 30 Guilford Street, London WC1N 1EH} \footnotemark[1] \\
    Great Ormond Street Institute of Child Health,\\
    University College London}
  \maketitle
} \fi

\if0\blind
{
  \bigskip
  \bigskip
  \bigskip
  \begin{center}
    {\LARGE\bf Modelling Network Time Series as Generalised Network Autoregressive Processes with Application to a COVID-19 Series.}
\end{center}
  \medskip
} \fi

\begin{abstract}
Network time series are becoming increasingly important across many areas in science and medicine
and are often characterised by a known or inferred underlying network structure, which can be exploited to make sense of dynamic phenomena that are often high-dimensional. 
For example, the Generalised Network Autoregressive (GNAR) models  exploit such structure parsimoniously.
We use the GNAR framework to introduce two association measures: the network and partial network autocorrelation functions,
and introduce  Corbit (correlation-orbit) plots for visualisation.
As with regular autocorrelation plots, Corbit plots permit
interpretation of underlying correlation structures and, crucially, aid model selection more rapidly than using
other tools such as AIC or BIC.
We additionally interpret GNAR processes as generalised graphical models, which constrain the processes' autoregressive structure and exhibit
interesting theoretical connections to graphical models via utilization of higher-order interactions. We demonstrate how 
incorporation of prior information is related to performing variable selection and shrinkage in the GNAR context. 
We illustrate the usefulness of the GNAR formulation, network autocorrelations and Corbit plots by modelling a COVID-19 network time series
of the number of admissions to mechanical ventilation beds at $140$ NHS Trusts in England \& Wales.
We introduce the Wagner plot that can analyse correlations over different time periods
or with respect to external covariates. 
In addition, we introduce plots that quantify the relevance and influence of individual nodes.
Our
modelling provides insight on the underlying dynamics of the COVID-19 series, highlights two groups of geographically
co-located `influential' NHS Trusts and demonstrates superior prediction abilities when compared to existing techniques.
\end{abstract}

\noindent%

\section{Introduction}
\label{sec:intro}
Network time series in many areas benefit from
the development of statistical methods that enable interpretation and inference of relationships present in dynamic phenomena. Modelling such network time series necessitates studying a constant flux of complex data characterised by a temporal component among large numbers of interacting variables.
The interacting variables are often associated with a network structure, for example, networks in
neuroscience, biology, medicine and business to name a few. In the absence of a network, a method can be
carried out to learn a possible network structure, which enables efficient modelling, analysis and forecasting
of a large number of interactions; see \cite{lauritzen, graph_lasso, Songsiri200989}. A complete overview of methods in network science and multivariate time series is beyond the scope of this work; see \cite{network_time_series_overview} for a review.
For a non-exhaustive overview of network methods in statistics {\em and} time series see, e.g.,
\cite{graph_series, lauritzen, luth, brockwell_davis, Songsiri200989, sna_book, ts_forecast_book} and
\cite{DALLAKYAN2022107557}.
\par
Recently, the generalized network autoregressive (GNAR) model has been developed
\cite{gnar_org, Zhu17, gnar_paper}, which provides a parsimonious interpretable model that has been often shown to
have both simpler interpretability {\em and} superior forecasting performance in a number of scenarios.
Model extensions in this rapidly
developing area include, for example, \cite{Zhu19} for quantiles, \cite{Zhou20} for Network GARCH models,
\cite{ArmillottaFokianos21} for Poisson/count data processes, 
\cite{NasonWei22} to admit time-changing covariate variables and \cite{Mantziou23} for GNAR processes on
the edges of networks.
 Such models have proven useful for many (network) time series where the characteristics of the series are similar from variable (node) to variable (node)
in a network, although GNAR's utility is not limited to this situation.

Crucial elements of any statistical modelling exercise are model elicitation and specification.
For many time series models, and for GNAR in particular, this involves
choosing
quantities such as the order $p, q$ of any autoregressive and moving average terms, respectively, and the differencing
parameter, $d$.
GNAR models also involve $p$,
and, in addition, the \mbox{$p\times 1$} vector, $s$:
the number of 
`stage-neighbours' per lag. Until now, the
main tools for GNAR model
order choice have been Akaike's and the Bayesian information criteria (AIC
and BIC),
GNAR versions
of which appear in the GNAR CRAN package developed by~\cite{gnar_package}
for R, for example.

In regular univariate time 
series modelling users benefit
from both AIC and BIC, but are
aware of their shortcomings,
especially for series that are not
long. For example,
several different models
possessing similar AIC/BIC values make it
difficult to choose between them.
In addition, computing AIC and BIC
can be time consuming for models with
geometric growth in parameters.
For example, for an order $p$
GNAR model
that admits $s$ stages of neighbours per lag
means estimating $s^p$ parameters.
Typically, the `final' model will
be parsimonious, but $s^p$ models might
need to be investigated with AIC/BIC
until that final model is found.
For example, for monthly data we
might want to start with $p=12$ to
enable detection of annual cycles and
with, e.g., $s=4$ this would result in
$4^{12} \approx 17$ million models to
be investigated.

For regular univariate
time series,
users additionally 
have access
the autocorrelation (acf) and partial
autocorrelation (pacf) functions,
which are well-known tools
that aid model order determination
and can also detect other behaviours
such as trend and seasonality.
Indeed, users often view acf and
pacf plots before any formal modelling.
A key contribution
of our work introduces
a network-enabled version of these tools:
Section~\ref{sec:corbit} introduces
the network autocorrelation function
(NACF) and the partial NACF (PNACF).
Section~\ref{sec:corbit} also introduces
a new graphical tool, the 
Corbit plot, to clearly convey
information in either
the NACF or PNACF
for a network time series
(see Figure~\ref{fig: corbit sim} below for
a preview).
Corbit plots enable
users to quickly and
directly help identify model
order and other behaviours for
network time series, just
as the acf and pacf
plots do for 
univariate series.

Section~\ref{sec: graph model connection} proposes an interpretation of GNAR processes as generalised graphical models, which constrain the processes' autoregressive structure and provides proof of some interesting connections to graphical models by incorporating higher-order interactions. It also proposes an interpretation of how including prior information into our analysis is related to performing variable selection and shrinkage for GNAR models. 
In particular, we prove a new result
explaining the connection
between our multi-stage GNAR
neighbourhood
structure and a hierarchy imposed
on the process
inverse cross-spectrum matrix.
Section~\ref{sec: graph model connection} also
 shows  how classical graphical time
 series models are a
 special case of GNAR
 processes.
 Section~\ref{sec:corbit} also exhibits
 Corbit plots on simulated data,
 clearly showing their advantage for model
 interpretation and selection.

Section~\ref{sec: application} examines the problem of modelling, analysis and prediction of
the number of patients occupying mechanical ventilation beds during the COVID-19 pandemic in $140$ NHS Trusts in England \& Wales.
Corbit plots conveniently and rapidly give strong guidance as to choice of GNAR process order that suggests
a very parsimonious model.
We demonstrate the superiority of the
obtained GNAR models for prediction
compared to established time series
models. We further introduce an extension of the Corbit plot, named
the Wagner plot, which permits
analysts
to understand (i) the 
effect of covariates on the
network time series correlation
structure or (ii) how the
correlation structure can change
over different time periods.
The latter can provide a
clear and immediate indication
of nonstationarity, where it exists. For the COVID-19 mechanical  ventilation beds we find that a Wagner
plot can
show how the dynamics of the process change
during different waves of the pandemic. The Wagner plot is so named after the composer who wrote `The Ring Cycle' and
the plot is composed of rings of circles. We also present two new plots that show the local influence and global relevance
of individual Trusts within the network. 

GNAR is an alternative method, which is useful when the data satisfy certain conditions, which we make explicit in Sections \ref{sec:gnar methodology} and \ref{sec: graph model connection}. Thus, GNAR should not be thought of as a general method for multivariate time series but rather as an addition to the existing toolbox. 

Next, Section~\ref{sec:gnar methodology} formulates a hierarchical representation for GNAR processes, which allows us to write the model in compact matrix notation and later efficiently
define the GNAR NACF and PNACF and associated Corbit and Wagner plots.

\section{GNAR Model and Methods}
\label{sec:gnar methodology}
A network time series $\mathcal{X} := (\boldsymbol{X}_t, \mathcal{G})$ is a stochastic process composed of a multivariate time series $\boldsymbol{X}_t \in \mathbb{R}^d$ and an underlying network $\mathcal{G} = (\mathcal{K}, \mathcal{E})$, where $\mathcal{K} = \{1, \dots, d\}$ is the set of nodes, $\mathcal{E}  \subseteq \mathcal{K} \times \mathcal{K}$ is the set of edges, and $\mathcal{G}$ is an undirected graph, which has $d$ nodes. Each univariate time series $X_{i, t} \in \mathbb{R}$ is associated to node $i \in \mathcal{K}$ in $\mathcal{G}$.
\par 
We review the global-$\alpha$ GNAR model for analysing network time series given our particular focus on self-similar interactions across nodes in the network. Also, throughout this work we assume that the network is static, however, we note that GNAR processes can handle time varying networks; see \cite{gnar_org, gnar_paper}. Before we present the model and estimation method we introduce two matrices that aid us in expressing the GNAR model in a more compact form.
\subsection{Weights and \texorpdfstring{$r$}{r}-stage adjacency matrices}
\begin{figure}
        \centering
        \begin{subfigure}{0.48\textwidth}
            \includegraphics[width=1.2\textwidth]{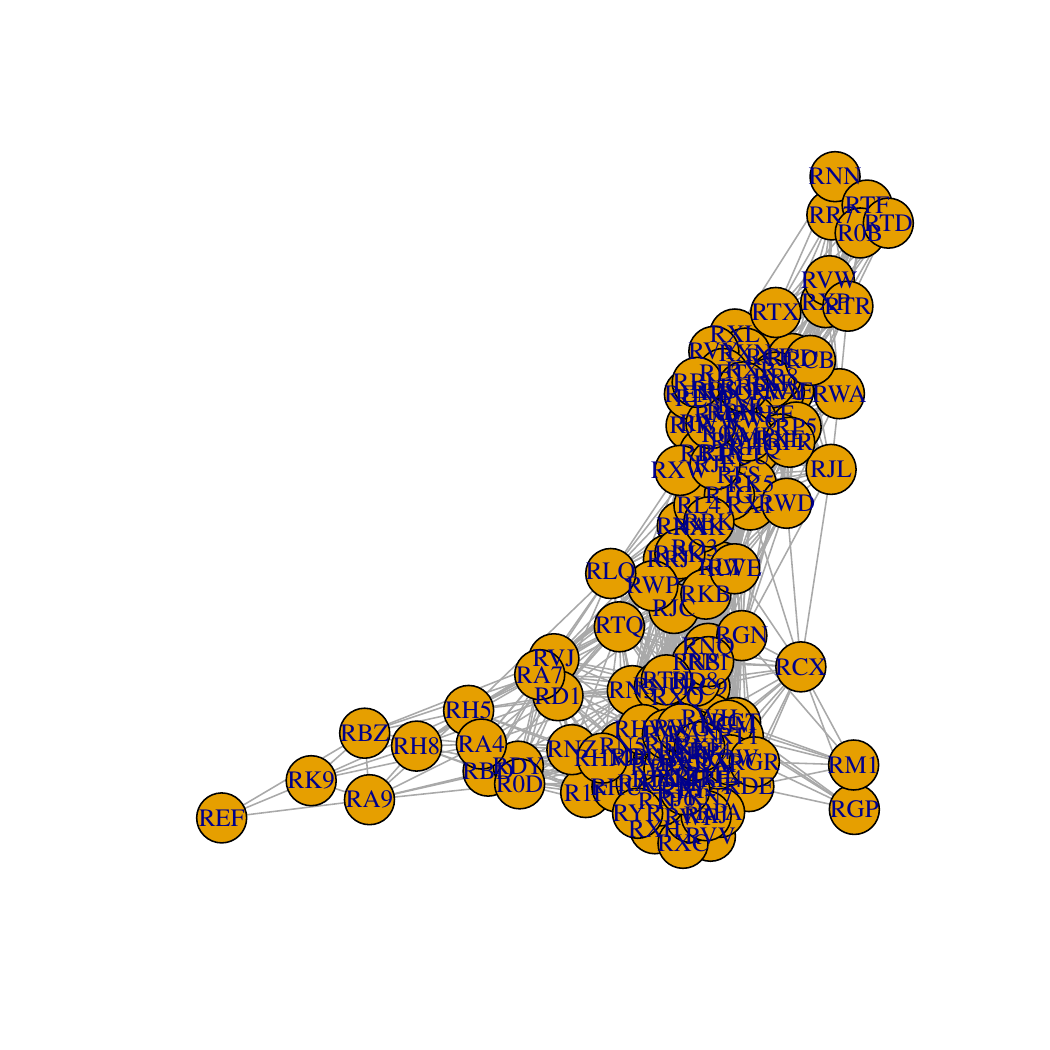}
            \caption{NHS Trusts network 1-stage neighbours.}
            \label{fig: 1 stage neighbors}
        \end{subfigure}
        \hfill
        \begin{subfigure}{0.48\textwidth}
            \includegraphics[width=\textwidth]{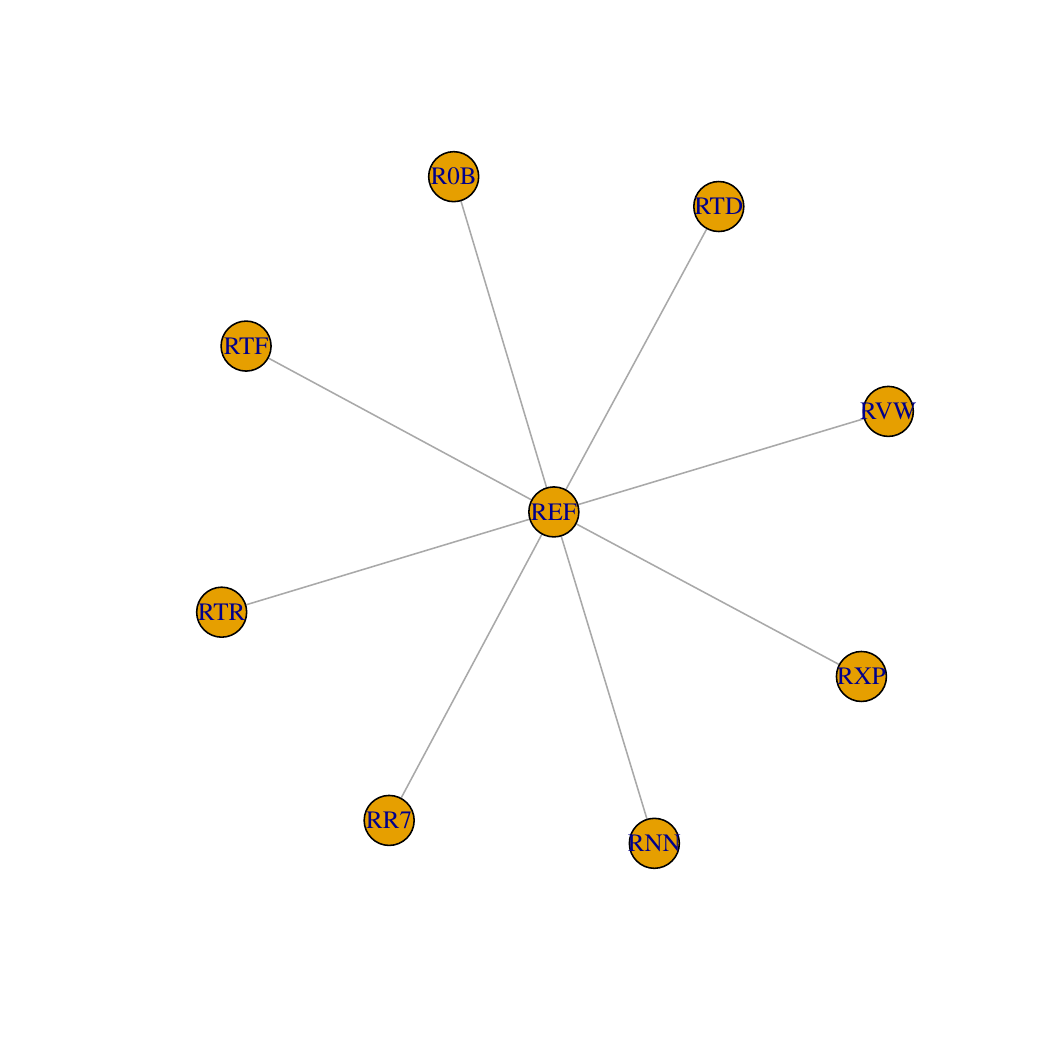}
            \caption{NHS Trusts network 6-stage neighbours.}
            \label{fig: 6 stage neighbors}
        \end{subfigure}
        \caption{(a) network with directly connected nodes. (b) network corresponding to 6-stage neighbours. NHS Foundation Trust Codes: REF=Royal Cornwall Hospitals;
        RVW=North Tees \& Hartlepool; RXP=County Durham \& Darlington;
        RNN=North Cumbria Integrated Care; RR7=Gateshead Health;
        RTR=South Tees Hospitals; RTF=Northumbria Healthcare;
        R0B=South Tyneside \& Sunderland; RTD=Newcastle upon Tyne Hospitals.}
        \label{fig: rstage neighbors}
    \end{figure}

GNAR processes propose a parsimonious model by exploiting the network structure. A key notion is that of $\rstage$ neighbours, we say that nodes $i$ and $j$ are $\rstage$ neighbours if and only if the shortest path on the network $\mathcal{G}$ between them has 'distance' equal to $r$, in the sense that the number of edges on the shortest path is equal to $r$, and define $\mathcal{N}_r (i) \subseteq \mathcal{K}$ as the set of $\rstage$ neighbours of node $i$. We account for this higher-order structure by introducing $\bmat{S}_r$, which is the $\rstage$ adjacency matrix.
\begin{definition} \label{def: rstage adjacency}
    Let $\mathcal{G} = (\mathcal{K}, \mathcal{E})$ be a network and let $\boldsymbol{X}_{t}$ be the network time series with underlying network $\mathcal{G}$, define the \mbox{$r$-stage} adjacency matrix as the matrix $\mathbf{S}_r$ $\in \mathbb{R}^{d \times d}$ with entries $[\mathbf{S}_r]_{ij}$, where each entry $[\mathbf{S}_r]_{ij} = 1$ if and only if node $i$ is an \mbox{$r$-stage} neighbour of node $j$, otherwise $[\mathbf{S}_r]_{ij} = 0$.
\end{definition}
Note that $\bmat{S}_1$ is the ordinary adjacency matrix and that nodes cannot be $\rstage$ neighbours for different choices of $r$. We illustrate this object by plotting the set of $\rstage$ neighbours for $r = 1, 6$ present in the network $\mathcal{G}$ associated to the COVID-19 (network) time series in Figure~\ref{fig: rstage neighbors}. For this data set, there are no neighbours at seven stages or higher.
The set of $\rstage$ neighbours of node $i$ can be found by looking at the $i$th row of $\bmat{S}_r$, furthermore, each $\bmat{S}_r$ is a symmetric matrix that can be computed sequentially from previous $\rstage$ adjacency matrices; see the supplementary material for a more thorough exposition. 
\par
We also consider the weights between nodes $i$ and $j$, each weight $w_{ij} \in [0, 1]$ quantifies the relevance node $j$ has on node $i$ with respect to neighbourhood regression. If the network $\mathcal{G}$ does not have weighted edges, a GNAR model assigns equal importance to each node $j$ in the set of $\rstage$ neighbours in the sense that if node $j$ is an $\rstage$ neighbour of node $i$, then $w_{ij} = \{ | \mathcal{N}_{r} (i) | \}^{-1} $. 
\par
If $\mathcal{G}$ has weighted edges $\tilde{w}_{ij}$, then each weight is normalised, so that the weights associated to each $\rstage$ neighbourhood add up to one, as follows $w_{ij} = \tilde{w}_{ij} \{ \sum_{l \in \mathcal{N}_r (i)} \tilde{w}_{il} \}^{-1}$. Hence, $\sum_{j \in \mathcal{N}_r (i)} w_{ij} = 1$ for all $\rstage$ neighbourhood sets. The connection between these weights and inverse distances is explored in Section \ref{sub: subsect oracle selection}.

A GNAR model performs autoregression for each nodal time series as well as neighbourhood regression. By neighbourhood regression we mean that for each nodal time series $X_{i,t}$ we compute its autoregression at lag $k$ with a convex linear combination $Z_{i, t}^{r}$ of its $\rstage$ neighbours, which for a fixed $r \in \{1, \dots, r_{\max}\}$ is given by
\begin{equation} \label{eq: node neigh regression}
    Z_{i, t}^{r} := \sum_{j \in \mathcal{N}_r (i)} w_{ij} X_{j, t},
\end{equation}
where $r_{\max} \in \mathbb{N}$ is the longest shortest path in the network $\mathcal{G}$ - i.e., $d(i, j) \leq r_{\max}$ for all node pairs $(i, j)$.
Since node $j$ can only be part of one neighbourhood regression with respect to all other nodes in $\mathcal{G}$ we can identify the unique connection weight between nodes $i$ and $j$ as $w_{ij}$ for all nodes in $\mathcal{G}$. To help express neighbourhood regression in matrix notation we introduce the weights matrix $\bmat{W}$ for GNAR models. 
\begin{definition} \label{def: gnar weight matrix}
     Let $\mathcal{G} = (\mathcal{K}, \mathcal{E})$ be a network and let $\boldsymbol{X}_{t}$ be the network time series with underlying network $\mathcal{G}$, the weights matrix is the matrix $\mathbf{W}$ $\in \mathbb{R}^{d \times d}$ with entries $[\mathbf{W}]_{ij} := w_{ij}$.
\end{definition}
Note that the diagonal entries are equal to zero because there are no self-loops in $\mathcal{G}$, and that $\bmat{W}$ is not necessarily symmetric because nodes can have different degrees of relevance across the network.
\subsection{GNAR Model}
By using the definitions above, we can express GNAR in matrix notation. Let $\odot$ denote the Hadamard (component-wise) product, which paired with $\bmat{S}_r$ and $\bmat{W}$ can be used to select the set of $\rstage$ neighbours for each node $i$ and compute the corresponding neighbourhood regression $Z_{i, t}^{r}$. The node-wise representation of a global-$\alpha$ GNAR model with maximum lag equal to $p \in \mathbb{N}$ and maximum $\rstage$ depth $s_k \in \{1, \dots, r_{\max} \}$ at each lag is given by
\begin{equation} \label{eq: node-wise}
    X_{i, t} = \sum_{k = 1}^{p} \big ( \alpha_k X_{i, t - k} + \sum_{r = 1}^{s_k} \beta_{kr} Z_{i, t - k}^{r} \big ) + u_{i, t},
\end{equation}
where the $\alpha_k \in \mathbb{R}$ are 'standard' autoregressive parameters and the $\beta_{kr} \in \mathbb{R}$ are neighbourhood autoregressive parameters for $r = 1, \dots, s_k$ at each lag $k = 1, \dots, p$. We denote this model order by $\GNAR (p, [s_1, \dots, s_p] )$, hence there are $p$ autoregressive terms and for each one of these there are $s_k$ neighbourhood regression terms which are given by (\ref{eq: node neigh regression}). Also, we assume that the $u_{i, t}$ are independent and identically distributed (IID) white noise with mean zero and variance $\sigma^2 > 0$ for all nodes. The model given by (\ref{eq: node-wise}) is an identical more compact representation  of the model in \cite{gnar_org}.
\par
Before expressing the model in (\ref{eq: node-wise}) by a vector-wise representation we introduce the $\rstage$ linear regression vector time series 
\begin{equation} \label{eq: vector projection}
    \boldsymbol{Z}_{t}^{r} := (\bmat{W} \odot \bmat{S}_r) \boldsymbol{X}_{t}.
\end{equation}
\par
Each $i$th entry in $\boldsymbol{Z}_{t}^{r}$ is equal to the $\rstage$ neighbourhood regression for node $i$ given by (\ref{eq: node neigh regression}). Now we can write the matrix notation version of the model in (\ref{eq: node-wise}) as
\begin{equation} \label{eq: vector wise}
    \boldsymbol{X}_{t} = \sum_{k = 1}^{p} \big ( \alpha_k \boldsymbol{X}_{t - k} + \sum_{r = 1}^{s_k} \beta_{kr} \boldsymbol{Z}_{t - k}^{r} \big ) 
    + \boldsymbol{u}_{t}, 
\end{equation}
where the $\alpha_k \in \mathbb{R}$ and $\beta_{kr} \in \mathbb{R}$ are the autoregressive coefficients in (\ref{eq: node-wise}) and $\boldsymbol{u}_t$ are (IID) multivariate white noise with mean zero and covariance matrix $\sigma^2 \bmat{I_d}$. The representation in (\ref{eq: vector wise}) highlights the parsimonious structure of a global-$\alpha$ GNAR model. Note that the number of parameters in the model increases not with the dimension of $\boldsymbol{X}_t$ and lag but rather with the depth of $\rstage$ regression and maximum lag. 
\par
A close look at (\ref{eq: vector wise}) reveals that the \mbox{global-$\alpha$} GNAR model can be written as a constrained VAR model; see \cite{brockwell_davis}
, for which the autoregressive matrices 
$$\bmat{\Phi}_k := \{\alpha_k \bmat{I}_d + \sum_{r = 1}^{s_k} \beta_{kr} (\bmat{W} \odot \bmat{S}_r ) \}, $$
are restricted by the network structure. \cite{gnar_paper} exploit this connection to show a more general result with respect to stationarity conditions for GNAR processes with a static network than the one we give below.

\begin{theorem}{[\textbf{\cite{gnar_paper}}]} \label{th: stationary cond}
Let $\boldsymbol{X}_t$ be a \mbox{global-$\alpha$} $\GNAR (p, [s_1, \dots, s_p] )$ process with associated static network $\mathcal{G} = (\mathcal{K}, \mathcal{E})$. If the autoregressive coefficients in (\ref{eq: vector wise}) satisfy 
$$
    \sum_{k = 1}^{p} ( |\alpha_k| + \sum_{r = 1}^{s_k} |\beta_{kr} | ) < 1, 
$$
then $\boldsymbol{X}_t$ is stationary.
\end{theorem}
We end this subsection by highlighting that GNAR models are special, highly parsimonious VAR models with parameter constrains informed by the underlying network structure. If a GNAR formulation is appropriate for modelling a particular data set, then it can be an extremely powerful forecasting tool, which enables interpretation of a large number of interactions. We describe these properties in Section \ref{sec: graph model connection} and illustrate its advantages by analysing the COVID-19 (network) time series in Section \ref{sec: application}.
\subsection{GNAR Model Estimation}
A global-$\alpha$ $\GNAR (p, [s_1, \dots, s_p] )$ model has $p$ autoregressive $\alpha_k$ coefficients and $\sum_{k = 1}^{p} s_k$ neighbourhood regression $\beta_{kr} $ coefficients. For comparison, a VAR($p$) model has $p d^2$ parameters, so as long as $q := p + \sum_{k = 1}^{p} s_k < p d^2$ the GNAR model given by (\ref{eq: vector wise}) needs to estimate far fewer parameters, this is particularly important for settings in which $p d^2$ is larger than the number of time step realisations observed, such as the COVID-19 data we analyse below.
\par
Assume that $T \in \mathbb{N}$ time steps of data $\bmat{X} := [\boldsymbol{X}_{1}, \dots, \boldsymbol{X}_{T} ]$ arising from a GNAR process with known order become available. Our objective is to estimate the unknown autoregressive coefficients $\alpha_k$ and $\beta_{kr}$ after fixing the lag-depth pair $(p, [s_1, \dots, s_k] )$. 
\par
To do this we assume that $\boldsymbol{X}_{t}$ is a stationary $\GNAR (p, [s_1, \dots, s_p] )$ process, and that $n = T - p$ is the number of observations for which there are $p$ previous observed lags available for estimation, and for $t = p + 1, \dots, T$ define the following.
\begin{align} \label{eq: estimation build-up}
    \boldsymbol{y}_t &:= \boldsymbol{X}_{p+1}, \nonumber \\
    \bmat{Z}^{1:s_1}_{t -1} &:= [\boldsymbol{Z}_{t - 1}^{1}, \dots, \boldsymbol{Z}_{t - 1}^{s_1}], \nonumber \\
    \bmat{R}_{t} &:= [\boldsymbol{X}_{t - 1} , \bmat{Z}^{1:s_1}_{t -1}, \dots, \boldsymbol{X}_{t - p} , \bmat{Z}^{1:s_p}_{t -p}],
\end{align}
where $\boldsymbol{y}_t \in \mathbb{R}^{d}$ is the data vector of 'responses', $\bmat{R}_t \in \mathbb{R}^{d \times q}$ is the design matrix at time-step $t$, and $\boldsymbol{Z}_{t - 1}^{s_1}$ are given by (\ref{eq: vector projection}). Furthermore, we define the vector of parameters $\boldsymbol{\theta} \in \mathbb{R}^q$ as
$$\boldsymbol{\theta} := (\alpha_1, \beta_{11}, \dots, \beta_{1 s_1}, \alpha_2, \dots, \beta_{p s_p}), $$
which is the vector of unknown linear coefficients.
\par
Thus, if $\boldsymbol{\theta}$ satisfies the assumptions in Theorem \ref{th: stationary cond}, the realisations
$\boldsymbol{y}_t = \bmat{R}_{t} \boldsymbol{\theta} + \boldsymbol{u}_{t}$ are statistically uncorrelated observations of linear models given by \eqref{eq: vector wise}, where $\boldsymbol{u}_t$ are the same as in (\ref{eq: vector wise}) for $t = p + 1, \dots, T$. Next, by concatenating the column vectors $\boldsymbol{y}_{t}$ into one column vector $\boldsymbol{y} \in \mathbb{R}^{nd}$ as well as the design matrices $\bmat{R}_{t}$ into one design matrix $\bmat{R} \in \mathbb{R}^{nd \times q}$ we can expand (\ref{eq: vector wise}) as the linear model
\begin{equation} \label{eq: T steps linear model}
    \boldsymbol{y} = \bmat{R} \boldsymbol{\theta} + \boldsymbol{u}.
\end{equation}
Now, we can couple the $n$ linear regression problems as if we had $n$ independent samples of size $d$ and fit the linear model by ordinary least-squares for data with sample size equal to $nd$ and $q$ unknown parameters, thus, the least-squares estimator for a $\GNAR (p, [s_1, \dots, s_p] )$ model with a \mbox{global-$\alpha$} specification is given by
\begin{equation} \label{eq: ls estimator}
    \hat{\boldsymbol{\theta}} = \big ( \myprod{\bmat{R}}{\bmat{R}} \big )^{-1} \myprod{\bmat{R}}{\boldsymbol{y}}, 
\end{equation}
where $\bmat{R}$ and $\boldsymbol{y}$ are given by (\ref{eq: estimation build-up}). 
\par
Note that if we further assume that the $\boldsymbol{u}_t \sim N(0, \sigma^2 \bmat{I_d})$, then $\Hat{\boldsymbol{\theta}}$ given by (\ref{eq: ls estimator}) is also the maximum likelihood estimator. We also point out that assuming that all the nodal white noise processes have the same variance might not be sensible, in that case it is possible to adapt (\ref{eq: T steps linear model}) into a generalised least-squares problem as well as other relaxations. Nevertheless, throughout this work we maintain the assumptions in Theorem \ref{th: stationary cond} and (\ref{eq: vector wise}), and estimate the unknown coefficients with $\Hat{\boldsymbol{\theta}}$ given by (\ref{eq: ls estimator}).

\section{Graphical Aids for Model Selection}
\label{sec:corbit}
In accordance with classical time series methodology we study the autocorrelation structure for different choices of maximum $\hlag$ and $\rstage$ depth, which we denote by $(h, r)$ throughout this section. With this goal in mind, we introduce the network autocorrelation function (NACF) for GNAR processes as well as the correlation-orbit (Corbit) plot as an effective graphical aid for performing model selection. The NACF allows us to quantify the correlation observed in the network with respect to the pair $(h, r)$. The Corbit plot enables us to efficiently visualise the correlation structure in the data if there is one, which is useful for model selection; see \cite{brockwell_davis}.
\par
The NACF name we choose coincides with the one in the \texttt{SNA} package aimed at social network analysis; 
see \cite{sna}. 
However, our NACF is targeted at network time series data and incorporates the weights and neighbourhood structure. We extend the notion of an autocorrelation function for a univariate time series by exploiting the structure the weights and $\rstage$ neighbourhoods give us by incorporating them in the NACF definition.
\subsection{GNAR Network Autocorrelation Function}
Examination of the NACF suggests the order of a GNAR model assuming that $\boldsymbol{X}_t$ can be modelled by (\ref{eq: vector wise}) and satisfies the conditions in Theorem \ref{th: stationary cond}. By comparing the NACF values for different choices of $(h, r)$ we can analyse the lag and $\rstage$ depth at which autocorrelation starts to decay. This permits us to study the autocorrelation for the entire network time series and avoids comparing all the cross-correlations for each pair of variables as would be required in a vector autoregression (VAR) model. Our NACF is defined below.
\begin{definition}\label{def: nacf}
With the same notation and definitions as above for $\bmat{W}$ and $\bmat{S}_r$, the network autocorrelation function of a $\GNAR$ process $\boldsymbol{X}_t$, with autocovariance bound
\newline
$\lambda := \bigg [ \underset{j = 1, \dots, d}{\max} \big \{\sum_{i = 1}^d [(\bmat{W} \odot \bmat{W})]_{i j} \big \} \bigg ]^{\frac{1}{2}},$ is given by
\begin{equation} \label{eq:nacf definition}
    \nacf(h, r) := \frac{\sum_{t=1}^{T - h} ( \boldsymbol{X}_{t + h} - \boldsymbol{\overline{X}})^{T} \big ( \mathbf{W} \odot \bmat{S}_r + \bmat{I_d} \big )
    ( \boldsymbol{X}_{t} - \boldsymbol{\overline{X}})}
    {\sum_{t=1}^{T}  ( \boldsymbol{X}_{t} - \boldsymbol{\overline{X}})^{T} \big \{ \big (1 + \lambda \big) \bmat{I_d} \big \} ( \boldsymbol{X}_{t} - \boldsymbol{\overline{X}})}.
\end{equation}
\end{definition}
We remark the following result, which connects the NACF to the ACF.
\begin{remark}
    If we model a univariate time series $X_t \in \mathbb{R}$ as a $\GNAR$ process with a one node graph, then the $\mathrm{NACF}$ given by Definition \ref{def: nacf} simplifies to 
    \begin{equation*}
        \nacf(h, r) = \frac{\sum_{t =1}^{T - h} (X_{t + h} - \overline{X}) (X_t - \overline{X}) }{\sum_{t =1}^{T} (X_{t} - \overline{X})^2},
    \end{equation*}
    which is the autocorrelation function from univariate time series analysis.
\end{remark}
Intuitively the NACF borrows strength from the network structure for computing the autocorrelation at an $\hlag$ and $\rstage$ pair. It treats $\boldsymbol{X}_t$ as a composite object and computes the autocovariance between each $X_{i, t}$ with itself $X_{i, t - h}$ and the lagged regression $Z_{i, t - h}^{r}$ of its $\rstage$ neighbours after centring both vectors around the empirical mean. Then it computes the ratio of the sum of these autocovariances and the sum of upper bounded autocovariances that result from the network constraints. For that reason, we call the parameter $\lambda$ in Definition \ref{def: nacf} the autocovariance bound specified by the network $\mathcal{G}$ and weights matrix $\bmat{W}$, which is necessary for ensuring that $-1 \leq \nacf(h,r) \leq 1$ for all possible choices of $(h, r)$. We explore a possible NACF interpretation in Section \ref{sub: nacf interpretation}; see the supplementary material for the NACF derivation and some of its properties. 
\subsection{Corbit Plot}
We introduce the Corbit plot by studying realisations coming from a stationary \mbox{global-$\alpha$} $\GNAR(2, [1, 1])$ process. Assume that we do not know the model order, we can study the network autocorrelation decay by plotting the observed NACF values via the Corbit plot.
\begin{figure}[H]
    \centering
    \includegraphics[scale=0.40]{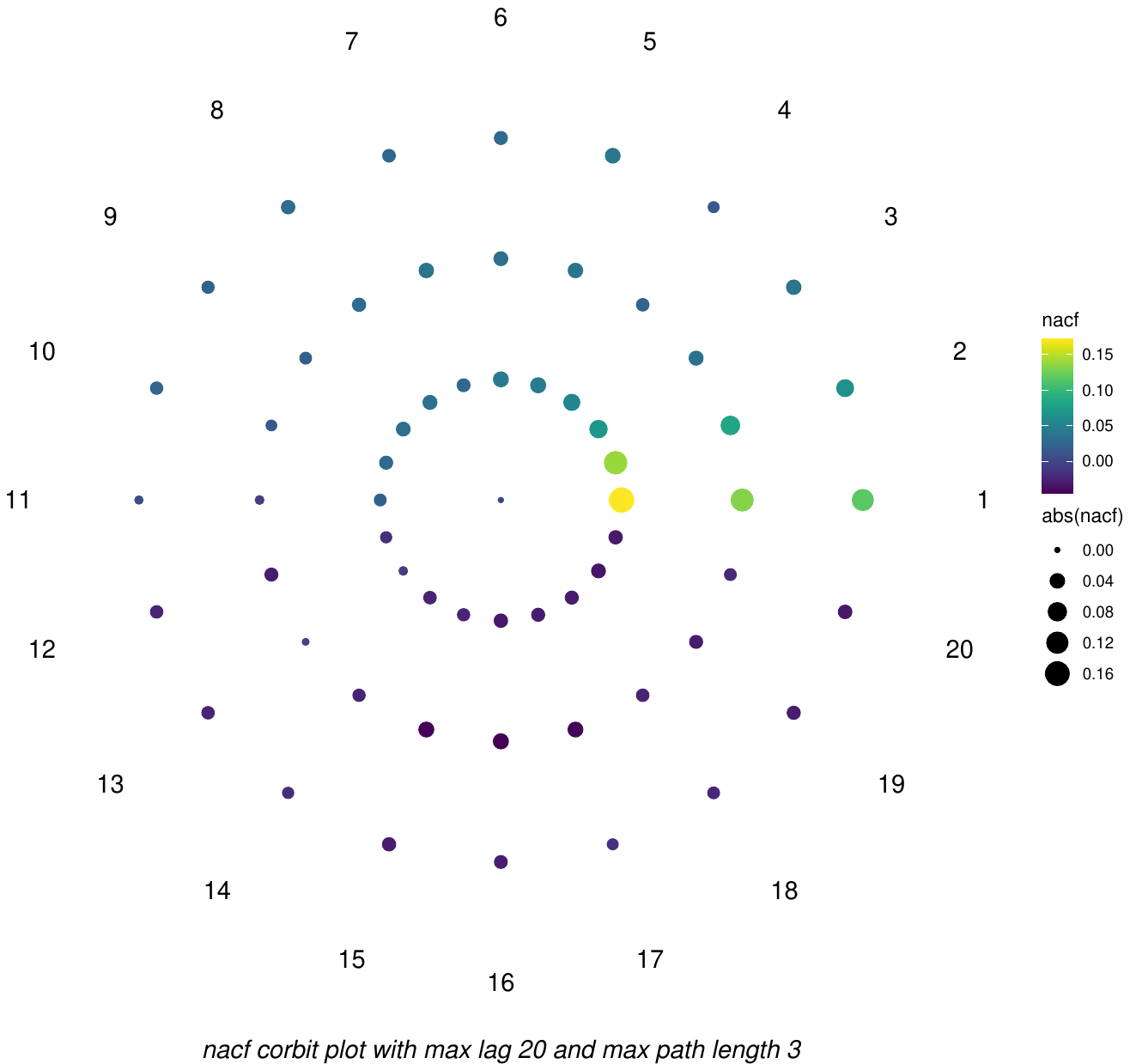}
    \caption{Corbit plot for 200 realizations from a stationary \mbox{global-$\alpha$} $\GNAR(2, [1, 1])$, where the underlying network is the fiveNet network included in the \texttt{GNAR} package; see \cite{gnar_package}. See text for description.}
    \label{fig: corbit sim}
\end{figure}
Figure \ref{fig: corbit sim} introduces the Corbit plot, each point corresponds to a \mbox{$h$-lag} and \mbox{$r$-stage} pair, and the colour is set by a colour scale based on the overall NACF values. The first ring depicts $1$-stage neighbours (i.e., nodes with one edge between them), the second ring considers \mbox{2-stage} neighbours (i.e., nodes with a shortest path length equal to two). In short, the ring number starting from the inside corresponds to $\rstage$ depth. 
\par
The numbers on the outside ring indicate the time lag used for computing the NACF. Therefore, the value of each point is $\nacf(h, r)$ where $h$ is the lag denoted on the the last ring and $r$ is the ring that corresponds to \mbox{$r$-stage} adjacency. 
\par
The point on the first inner ring with a number one to the right in Figure \ref{fig: corbit sim} is the value of $\nacf(1, 1)$, the second point on the first inner ring is $\nacf(2, 1)$, and the first point on the second inner ring is $\nacf(1, 2)$; this pattern repeats for subsequent rings and lags.
\par
Currently, in our software, there are two possible ways for assigning size to each point. The default choice is to use the NACF absolute value - i.e., $|\nacf(h, r)|$. This choice highlights the different $\nacf(h, r)$ magnitudes. An alternative choice is based on the $\GNAR$ conditional mean fixing lag and stage produces; see the supplementary material. 
\par
Finally, the point at the centre has NACF value equal to zero and the smallest size, which highlights the larger NACF values and facilitates comparing the NACF values to $\nacf(h, r) = 0$. We note that these choices are not exclusive and other measures of model fit and/or correlation could be used for assigning point size and colour. 
\par
Corbit plots are produced using the \texttt{viridis} R library, which provides a colour scale that is easily perceived by viewers with common forms of colour blindness (\cite{viridis_lib}) and the \texttt{ggplot} package functionality (\cite{ggplot_lib}). 
\par
The Corbit plot in Figure \ref{fig: corbit sim} shows that the NACF decays on/after lags equal to or larger than four across all stages. Also, the NACF drops after the second stage at the first lag and after the first stage for the second lag. Observe that the further separated nodes are in the network the closer the NACF is to zero. Furthermore, across all $\rstage$ depths we see that the NACF decays to zero as the lag increases. This Corbit plot suggests that autocorrelation for the simulated network time series has larger values for $h \in \{1, 2\}$ and $r \in \{1, 2, 3\}$ and decays sharply as the $\hlag$ and $\rstage$ depth increase. 
\par
This is in accordance with univariate ACF plots which decay as the lag increases, and does reflect the known underlying $\GNAR(2, [1, 1])$ structure.
\subsection{GNAR Partial Autocorrelation Function}
The NACF computes the autocorrelation between $\boldsymbol{X}_t$ and its lagged observations for a specific choice of $\rstage$ neighbourhood regression, however, it does not account for the effects previous and intervening lags and/or $\rstage$ neighbours have when computing $\nacf(h, r)$. This makes diagnosing model order by looking at the NACF values on a Corbit plot challenging since we do not know if autocorrelation has not reduced because of the effects previous lags and/or $\rstage$ depths might have. This difficulty could prevent us from observing the underlying GNAR autocorrelation structure if there is one.
\par
Motivated by techniques from univariate time series analysis, we propose the partial network autocorrelation function (PNACF) as a tool for diagnosing GNAR model selection. The PNACF computes the autocorrelation between $\boldsymbol{X}_t$ and $h$-lagged observations of itself for a specific $\rstage$ neighbourhood regression after the linear effects of previous lags and $\rstage$ neighbours have been removed. The PNACF acts as the partial autocorrelation function does for univariate time series by identifying model order from examining the $(h, r)$ pair after which there is a sharp decline in autocorrelation.
\par
Assume that $\boldsymbol{X}_t$ is a GNAR model given by (\ref{eq: vector wise}) and satisfies the conditions in Theorem \ref{th: stationary cond}, then it is possible to remove the effects from previous lags and $\rstage$ neighbours by focusing on the empirical residuals arising from a $\GNAR(h - 1, [(r - 1), \dots, (r - 1)])$ fit. These residuals correspond to the best linear prediction restricted to lag $(h - 1)$ and maximum $\rstage$ depth $(r - 1)$, which we denote by $\boldsymbol{X}_{t}^{h - 1, r - 1}:= \sum_{k = 1}^{h - 1} ( \alpha_k \boldsymbol{X}_{t - k} + \sum_{s = 1}^{r - 1} \beta_{kr} \boldsymbol{Z}_{t - k}^{r - 1}  )$; see the supplementary material. 
\par
One possible extension of the PACF from a univariate setting to the GNAR framework is defining the PNACF as the NACF between the residuals arising from the best linear predictions using $(h - 1)$ lags and $(r - 1)$ $\rstage$ neighbourhood regressions for $\hlag$ and $\rstage$ pairs. Unfortunately, computing the PNACF as mentioned above requires us to have prior knowledge of the autoregressive coefficients in (\ref{eq: vector wise}). We circumvent this by inputting the least-squares estimators as if they were the true parameter values, which is valid given the consistency of the estimator given by (\ref{eq: ls estimator}); see \cite{gnar_paper}. The values we get using the imputed parameters should reflect the underlying GNAR autocorrelation structure if there is one. The new forecast value is $\hat{\boldsymbol{X}}_{t}^{h - 1, r - 1} := \sum_{k = 1}^{h - 1} ( \hat{\alpha}_k \boldsymbol{X}_{t - k} + \sum_{s = 1}^{r - 1} \hat{\beta}_{ks} \boldsymbol{Z}_{t - k}^{s}  )$. Our PNACF is defined below.
\begin{definition} \label{def: pnacf}
For a stationary $\GNAR$ process $\boldsymbol{X}_t$ compute the residuals coming from a $\GNAR(h - 1, [(r - 1)])$ fit, the corresponding residual mean $\overline{\boldsymbol{u}}$, and the residuals with $\hlag$s between them
    $\boldsymbol{\hat{u}}_{t + h} = \boldsymbol{X}_{t + h} - \hat{\boldsymbol{X}}_{t + h}^{h - 1, r - 1}$ and
    $\boldsymbol{\hat{u}}_{t} = \boldsymbol{X}_{t} - \hat{\boldsymbol{X}}_{t}^{h - 1, r - 1}.$
Then, the sample partial network autocorrelation function is given by
\begin{equation} \label{eq: pnacf}
    \pnacf(h, r) := \frac{\sum_{t=1}^{T - h} ( \boldsymbol{\hat{u}}_{t + h} - \boldsymbol{\overline{u}})^{T} \big ( \bmat{W} \odot \bmat{S}_r + \bmat{I_d} \big )
    ( \boldsymbol{\hat{u}}_{t} - \boldsymbol{\overline{u}})}
    {\sum_{t=1}^{T}  ( \boldsymbol{\hat{u}}_{t} - \boldsymbol{\overline{u}})^{T} \big \{ \big (1 + \lambda \big) \bmat{I_d} \big \} ( \boldsymbol{\hat{u}}_{t} - \boldsymbol{\overline{u}})}, 
\end{equation}
where $\lambda$ is the same as in Definition \ref{def: nacf}.
\end{definition}
The PNACF computes the remaining network autocorrelation between residuals after removing the linear effects of previous lags and stages. Intuitively, it will cut-off to zero at every $(h, r)$ pair where $h > p$ and $r > r^*$, where $r^* := \max \{ s_1, \dots, s_p \}$ is the largest active $\rstage$ depth, since these pairs correspond to the sum of cross-correlations between the IID white noise processes $u_{i, t + h}$ and $u_{j, t}$ given by (\ref{eq: node-wise}).
By computing said network autocorrelations, the PNACF highlights the stage and lag at which the network autocorrelation cuts-off.
\begin{figure}[H]
    \centering
    \includegraphics[scale=0.45]{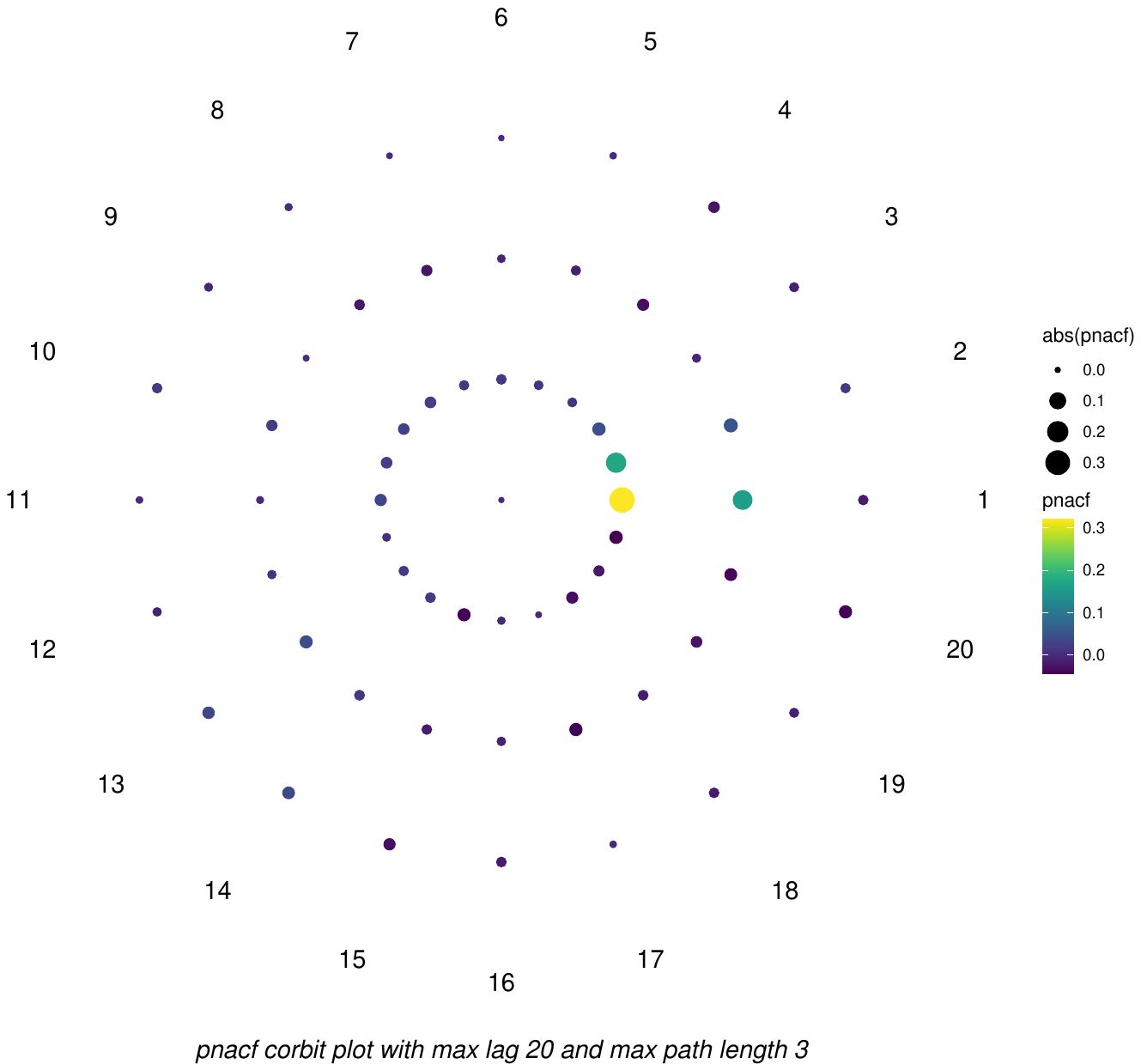}
    \caption{Corbit plot for 200 realizations from a stationary \mbox{global-$\alpha$} $\GNAR(2, [2, 1])$ which has the fiveNet network as underlying structure; see \cite{gnar_package}. The maximum lag is equal to 20 and maximum $\rstage$ is equal to 3.}
    \label{fig: pnacf sim}
\end{figure}
Figure \ref{fig: pnacf sim} shows that the PNACF cuts-off at lag three across all stages (i.e., it cuts-off for all $\rstage$ depths for $h \geq 3$), and that it cuts-off at stage one for lag two and at stage two for lag one. This Corbit plot suggests fitting a $\GNAR(2, [2, 1])$ which recovers the known data-generating process in this case. Note that the PNACF cut-off mimics the way in which the PACF decays when looking at univariate time series; see \cite{brockwell_davis}. We will introduce the Wagner plot below,
which is a development of the Corbit plot that highlights non-stationarities or show the effect of covariates. 

\section{GNAR Properties and Useful Interpretations}
\label{sec: graph model connection}
GNAR processes generalise graphical models for multivariate time series by introducing higher-order interactions between nodes, and propose a parsimonious specification that leverages the similarities of the individual node-wise processes.
For our COVID pandemic analysis later it is reasonable to expect that the SARS-CoV-2 virus
behaves similarly in different locations across the U.K. This section makes
the connection between GNAR and graphical models for multivariate time series explicit and 
further explains
the selection and shrinkage properties that GNAR models implicitly possess due to their
incorporation of prior knowledge.
\subsection{Connection to Graphical Models for Time Series}
A goal of GNAR is to include possible interactions between nodes that are not directly connected in the network. One possible way of generalising the notions in \cite{graph_series} is to extend edge-based interactions by assigning $\rstage$ adjacency based on a cross-spectral hierarchy presented in this subsection. GNAR introduces higher-order interactions into the graphical model by allowing $\rstage$ neighbourhood regression for $r \in \{1, \dots, r^*\}$, where $r^*$ is the largest active $\rstage$ neighbourhood regression. 
\par
These higher-order interactions can be interpreted as weaker dependence between nodes the further separated the nodes are in the graph, in the sense that $\rstage$ neighbourhood regression is less influential the larger $r$ is and ultimately non-influential if $r > r^*$. Here, the idea of an edge between nodes is extended to membership in $\rstage$ adjacency sets. Denote by $\mathcal{N}_r$ the set of $\rstage$ neighbours, so if $j \in \mathcal{N}_r (i)$, then, by symmetry and shortest path uniqueness, $(i, j) \in \mathcal{N}_r$, moreover, note that $\mathcal{N}_1$ is the ordinary set of edges.

An intuition for GNAR is that the cross-correlation between $X_{i, t + h}$ and $X_{j, t}$ at all lags $h$ should be strongest if
$j \in \mathcal{N}_1 (i)$, drop for $j \in \mathcal{N}_r (i)$ where $r \in \{2, \dots, r^*\}$,  and that $X_{i, t+ h}$ and $X_{j, t}$ do not heavily influence each other if $j \notin \mathcal{N} (i)$, where $\mathcal{N} (i) := \cup_{r = 1}^{r^*} \mathcal{N}_r (i)$ is the borough of node $i$ (i.e., collection of neighbourhoods). This motivation relates $\rstage$ neighbourhood regression to the cross-spectrum and the inverse spectral matrix, as next.
\begin{theorem} \label{th: pop_sparsity gnar}
    Let $\boldsymbol{X}_t$ be a stationary $\GNAR(p, [s_1, \dots, s_p])$ process with a static network structure $\mathcal{G} = (\mathcal{K}, \mathcal{E})$, full rank spectral matrix $\bmat{f} (\omega)$ and maximum active $\rstage$ depth $r^*$, then the inverse spectral matrix $\bmat{S}(\omega)$ and the node-wise distances $d(i, j)$ computed on the network $\mathcal{G}$ satisfy
        \par
        $\text{a. }$ There exists a partial correlation graph $\tilde{\mathcal{G}} = (\mathcal{K}, \tilde{\mathcal{E}})$ with the same set of nodes as $\mathcal{G}$ such that
        $$(i, j) \notin \tilde{\mathcal{E}} \textit{ if and only if }  d(i, j) \geq 2 r^* + 1.$$
        \par
        $\text{b.}$ There exists a cross-spectral hierarchy $\xi^{(1)} > \cdots > \xi^{(r^*)} >  \xi^{(r^* + 1)} = 0$ and an active $\rstage$ neighbourhood regression which satisfy
         $$d(i, j) \in \{2r - 1, 2r\} \textit{ if and only if  } \xi^{(r)} \leq |[\bmat{S}(\omega)]_{i j}| < \xi^{(r - 1)},$$
         for all $\omega \in (-\pi, \pi]$ and for all $r \in \{1, \dots, r^*\}$.
\end{theorem}
See the supplementary materials for a proof of Theorem \ref{th: pop_sparsity gnar}, and \cite{brockwell_davis, ts_forecast_book} for definitions of $\bmat{S}(\omega)$ and $\bmat{f}(\omega)$. Theorem \ref{th: pop_sparsity gnar} shows that $X_{i,t}$ and $X_{j, t + h}$ are uncorrelated at all lags $h$ given all the other nodes if and only if nodes $i$ and $j$ do not have common active $\rstage$ neighbours. It extends the notion that nodes without edges between them in a graph are uncorrelated given all the other nodes in said graph, by proposing that if the distance between nodes $i$ and $j$ is larger than $2 r^* + 1$, then $X_{i, t}$ and $X_{j, t + h}$ are uncorrelated given all the other nodal time series at all lags, which we express as nodes $i$ and $j$ do not collide in $\rstage$ neighbourhood regression.
\par
Moreover, Theorem \ref{th: pop_sparsity gnar} exhibits that the higher-order autocorrelation structure a GNAR process induces is equivalent to the process having a hierarchical dependence structure, which can be identified from the inverse spectral matrix (i.e., $\bmat{S}(\omega)$ acts similarly as the concentration matrix does for Gaussian graphical models). This property allows us to interpret the relevance different nodes $j$ have with respect to node $i$ by looking for the $\rstage$ neighbourhood regressions at which $j$ is active (i.e., if $j \in \mathcal{N}_r (i)$, then the smaller $r$ is the more relevant $j$ is). Note that if we restrict GNAR to \mbox{$1$-stage} neighbourhood regression, then higher-order interactions are not considered and the GNAR induced correlation structure is equivalent to a graphical model for multivariate time series. In that case, Theorem \ref{th: pop_sparsity gnar} recovers results of \cite{graph_series}.
\subsection{Oracle Selection and Shrinkage} \label{sub: subsect oracle selection}
Prior knowledge of the network can be interpreted as having an oracle solution for selecting active nodes in each $\rstage$ neighbourhood regression. A GNAR formulation exploits this by proposing a reparameterization of a constrained VAR process in a parsimonious manner that is related to autoregressive coefficient shrinkage, which reduces estimator variance and improves model interpretability.

\subsubsection*{\textit{For Selection}}
Before performing selection, each node-wise autoregression includes all possible nodes as predictors and the equivalent non-constrained VAR($p$) model is
\begin{equation} \label{eq: var}
     X_{i, t} = \sum_{k = 1}^{p} \big( \phi_{k}^{ii} X_{i , t- k} + \underset{j \neq i}{\sum} \phi_{k}^{ij} X_{j , t- k}
    \big )+ u_{i, t},
\end{equation}
where $\phi_{k}^{ij}$, $\phi_{k}^{ii} \in \mathbb{R}$ are autoregressive coefficients and the $u_{i,t}$ are IID white noise.

Model~\eqref{eq: var} has $d$ non-zero coefficients for each node-wise autoregression at every lag, hence, the total number of unknown parameters is $p d^2$. Following \cite{statLearn}, we say that a node-wise regression is $m$-sparse if only a subset of $m$ predictor nodes have a non-zero autoregressive coefficient. By definition we have that $m \leq d$, however, our focus is on models which highlight a small subset of relevant node predictors (i.e., $m \ll d$). Next, we constrain the VAR model~\eqref{eq: var}
by assuming that only nodes which satisfy $d(i, j) \leq r^*$, where the distances are computed on the underlying network, have a non-zero coefficient. Essentially, for all lags $k \in \{1, \dots, p\}$ we impose the constraint
\begin{equation} \label{eq: var constrain}
    \phi_{k}^{ij} \neq 0 \textit{ if and only if } d(i, j) \leq r^*,
\end{equation}
by symmetry, we also have that $\phi_{k}^{ij} \neq 0$ if and only if $\phi_{k}^{ji} \neq 0$ at all lags $k$.
\par
Applying  constraint \eqref{eq: var constrain} to the VAR model~\eqref{eq: var} and noting that $j \in \mathcal{N}(i)$ if and only if $d(i, j) \leq r^*$ gives
\begin{equation} \label{eq: sparse var}
    X_{i, t} = \sum_{k = 1}^{p} \big( \phi_{k}^{ii} X_{i , t- k} + \underset{j \in \mathcal{N} (i)}{\sum} \phi_{k}^{ij} X_{j , t- k}
    \big )+ u_{i, t},
\end{equation}
where $\mathcal{N}(i) = \cup_{r = 1}^{r^*} \mathcal{N}_{r} (i)$ and the $u_{i,t}$ are IID white noise. Above, each $i$th nodal time series has at most $m^{(i)} := |\mathcal{N}(i)| + 1 \leq d$ non-zero autoregressive $\phi_{k}^{ij}$, $\phi_{k}^{ii}$ coefficients at each lag $k$. Thus, under  constraint~\eqref{eq: var constrain}, each node-wise regression is $m^{(i)}$-sparse at each lag and has at most $p m^{(i)}$ unknown parameters. 
\par
We set $m := \max \{ m^{(i)} \}$ and see that all node-wise autoregressions are $m$-sparse and that the model has at most $p m^2$ unknown parameters. Furthermore, constraint~\eqref{eq: var constrain}
performs variable selection for all node-wise autoregressions, and reflects
our assumption of closer nodes being more relevant to each other.

Variable selection can be performed by multiplying the vector time series by the sum of $\rstage$ adjacency matrices. Let $\bmat{S} = \sum_{r = 1}^{r^*} \bmat{S}_r$, and note that $[\bmat{S}]_{ij} \neq 0$ if and only if $d(i, j) \leq r^*$, and that $[\bmat{S}]_{ii} = 0$ for all nodes $i$, hence, model~\eqref{eq: sparse var} is equivalent to
\begin{equation} \label{eq: selection equivalence}
    X_{i, t} = \sum_{k = 1}^{p} \big( \phi_{k}^{ii} X_{i , t- k} + \underset{j = 1}{\sum^d} \phi_{k}^{ij} [\bmat{S}]_{ij} X_{j , t- k}
    \big )+ u_{i, t},
\end{equation}
which has the same active nodes (i.e., $\phi_k^{ij} [\bmat{S}]_{ij} \neq 0 \iff d(i, j) \leq r^*$) as the GNAR vector-wise representation~\eqref{eq: vector wise}. This establishes variable selection equivalence between a GNAR formulation and a constrained VAR, where the number of selected nodes for each node-wise autoregression satisfies $m^{(i)} = \sum_{j = 1}^{d} [\bmat{S}]_{ij}$.
\par
Moreover, fixing $r^*$ in a GNAR formulation is equivalent to imposing a $\ell_0$-ball constraint on the autoregressive coefficients, which can be efficiently approximated by an $\ell_1$-norm constraint; see \cite{wainwright_2019}. Further, decreasing $r^*$ reduces the number of nodes included in node-wise autoregressions, which results in a sparser model by noting that at all lags $k$
\begin{equation*}
    0 < \normx{ \bmat{\Phi}_k \odot \bmat{S}_1 }_1 \leq \nu^{(1)} \leq \cdots \leq  \normx{ \bmat{\Phi}_k \odot \bmat{S} }_1 \leq \nu <  \normx{ \bmat{\Phi}_k }_1,
\end{equation*}
hence, as $r^*$ decreases the number of active nodes decreases too, which results in a smaller $\nu > 0$ (i.e., a tighter constraint) and a sparser model.
\par
Based on the above, we interpret $\rstage$ neighbours as the relevant predictors each node has, and we can think of maximum $\rstage$ depth (i.e., $r^*$) as a hyperparameter that controls variable selection.
\subsubsection*{\textit{For Shrinkage}}
GNAR accounts for the possibility that nodes in the same $\mathcal{N}_r$ are likely to be correlated, which will affect estimator and predictive performance (under a valid GNAR model); see \cite{statLearn}. Motivated by this, a GNAR formulation performs shrinkage on the selected variables at the population level by reparametrizing the constrained VAR model as follows. 
\par
Using the notation and concepts from a GNAR formulation the node-wise autoregressions in a sparse VAR($p$) are given by (\ref{eq: sparse var}). We assume that relevance weights $\sigma_{ij} > 0$ between nodes are available, which quantify the similarity between nodes $i$ and $j$, and can be computed as $\sigma_{ij} := \{h(i,j)\}^{-1}$, where $h(i,j) > 0$ measures a notion of distance between nodes, which does not have to be equal to the distance in the graph (but could be),
for instance, the distance can be a function of some exogenous variable that measures `closeness' between nodes in a non-linear manner. Essentially, the $\sigma_{ij}$ are a form of hard prior knowledge.
\par
The notion of closeness and `similarity' between nodes that the $\sigma_{ij}$ provide can be incorporated into the GNAR framework by reparameterizing the autoregressive coefficients in (\ref{eq: sparse var}) as $\beta_{kr} \sigma_{ij} :=  \phi_{k}^{ij} [\bmat{S}]_{ij}$, where prior knowledge of $\sigma_{ij}$ makes the $\beta_{kr}$ identifiable. Also, it means that if $d(i, j) > r^*$, then $\beta_{kr} \sigma_{ij} = 0$. This parameterization assumes that there is a global effect $\beta_{kr}$ common across $\rstage$ neighbours, which results in a parsimonious representation, i.e.\ a GNAR
model is valid.
\par
To satisfy the assumptions of Theorem \ref{th: stationary cond} (i.e., weights less than or equal to one), we normalise the weights by computing updates as $w_{ij} := \sigma_{ij} ( \sum_{l \in \mathcal{N}_r (i)} \sigma_{il} )^{-1}$, hence, 
$\sum_{j \in \mathcal{N}_r (i)} w_{ij} = 1$ for all nodes at every active $\rstage$ neighbourhood regression. Finally, to transform (\ref{eq: sparse var}) into a model equivalent to a GNAR model shrink the active $\phi_{k}^{ij}$ coefficients as follows
\begin{equation} \label {eq: shrink gnar}
\gamma_{k}^{i j} :=  \phi_{k}^{i j} \{ 1 + \sigma_{i j}^{-1} v_{r}(i, j) \}^{-1},
\end{equation}
where $ v_r (i, j) := \sum_{l \in \mathcal{N}_r (i) - \{j \} } \sigma_{il}$ - i.e., weight normalisation can be thought of as a constraint on the $\ell_2$-norm of the active coefficients. We further note an interesting connection between the above and parameter updates in linear ridge regression.

\begin{proposition} \label{prop: shrinkgae relation}
Let $\boldsymbol{y} \in \mathbb{R}^n$ be a vector of responses, $\bmat{A} \in \mathbb{R}^{n \times m}$ the design matrix and $\boldsymbol{\phi} \in \mathbb{R}^m$ the vector of unknown linear coefficients. Next define $\hat{\boldsymbol{\phi}} := \argmin \{ || \boldsymbol{y} - \bmat{A} \boldsymbol{\phi} ||_2^2 \},$ and $\hat{\boldsymbol{\gamma}} := \argmin \{ || \boldsymbol{y} - \bmat{A} \boldsymbol{\gamma} ||_2^2 \}$ such that $|| \boldsymbol{\gamma}||_2 \leq \lambda$, where $\boldsymbol{\gamma} \in \mathbb{R}^m$ and $\lambda > 0$. If the design matrix admits the singular value decomposition $\bmat{A} = \bmat{U} \bmat{\Sigma} \bmat{V}^T$, then we have that
\begin{equation} \label{eq: shrink sparse ridge}
    \hat{\gamma}_j = \hat{\phi}_j (1 + v \sigma_j^{-2} )^{-1},
\end{equation}
where $\sigma_{j}^{2}$ are the diagonal entries in $\bmat{\Sigma}^2$ and $v > 0$ is the Lagrange multiplier linked to the constraint $|| \boldsymbol{\gamma}||_2 \leq \lambda$.
\end{proposition}

In view of Proposition \ref{prop: shrinkgae relation}, and by comparing (\ref{eq: shrink gnar}) with (\ref{eq: shrink sparse ridge}), we interpret each $i$th node-wise $\rstage$ neighbourhood regression as a sparse linear regression with shrunken coefficients, which satisfy the weight normalisation constraint - i.e., 
$\gamma_{k}^{ij} =  \phi_{k}^{i j} \{ 1 + \sigma_{i j}^{-1} v_{r}(i, j) \}^{-1} [\bmat{S}]_{ij}$ for all nodal pairs at all lags. Moreover, by (\ref{eq: shrink gnar}) and the above, 
the usual estimated GNAR coefficients can thus be re-expressed by a particular transformed and shrunk version of VAR coefficients
that we have just explicitly represented by $\gamma_{k}^{ij} = \beta_{kr} w_{ij}$.
\par
Above, the $\beta_{kr} w_{ij}$ are the node-wise coefficients in the GNAR parameterization given by ~\eqref{eq: node-wise}, which is a constrained reparametrization of the sparse VAR($p$) process~\eqref{eq: sparse var} with shrunken coefficients $\gamma_{k}^{ij}$. This reparametrization results in a parsimonious model with $p + \sum_{k = 1}^{p} s_k \ll p m^2 \ll p d^2$ unknown parameters. Hence, GNAR processes can be interpreted as constrained VAR models which exploit knowledge of the underlying network structure to perform variable selection and shrinkage at the population level. Our GNAR framework produces parsimonious models that are highly interpretable.

\subsection{NACF Interpretation}
\label{sub: nacf interpretation}
The connection between GNAR processes, graphical models for multivariate time series and sparse VAR models allows us to interpret the NACF as a measure of the constrained (i.e., network induced) correlation structure in the process. This effect is accounted for by the autocovariance bound $\lambda$ given by Definition \ref{def: nacf}.
\par
The sparseness and shrinkage properties GNAR models have permit us to bound the dot product
$\langle \boldsymbol{x}, ( \mathbf{W} \odot \bmat{S}_r + \bmat{I_d}) \tildesym{x} \rangle $ for all $\rstage$ neighbourhood regressions as follows
\begin{equation} \label{eq: nacf inequality}
    \big | \big \langle \boldsymbol{x}, ( \mathbf{W} \odot \bmat{S}_r + \bmat{I_d}) \tildesym{x}  \big \rangle \big | 
    \leq ( 1 + \lambda) \normx{\boldsymbol{x}} \normx{\tildesym{x}},
\end{equation}
which illustrates the inner workings of the NACF; see the supplementary material for a proof of (\ref{eq: nacf inequality}). Essentially, $\lambda$ is a constant depending on the network structure that globally takes into account and corrects for the selection and shrinkage properties the weights $w_{ij}$ and $\rstage$ adjacency have on the network time series.
\par
The underlying GNAR structure points vectors in the direction of $(\mathbf{W} \odot \bmat{S}_r + \bmat{I_d})$ for which the magnitudes of the cross-covariances (i.e., the centered dot products) between $\boldsymbol{X}_{t + h}$ and $( \mathbf{W} \odot \bmat{S}_r + \bmat{I_d})\boldsymbol{X}_{t}$ are bounded by the magnitude of the autocovariance between independent components being projected onto the hypersphere $\{ \boldsymbol{x}: \normx{\boldsymbol{x} - \overline{\boldsymbol{x}}}_2 \leq (1 + \lambda) \}$. 
\par
By Theorem \ref{th: pop_sparsity gnar} and Section \ref{sub: subsect oracle selection}, decreasing $r^*$ makes the model sparser and $\lambda$ smaller, the Corbit plot exploits this to highlight the lag and stage pairs at which the NACF cuts-off. Also, since the operator norm of $(\mathbf{W} \odot \bmat{S}_r + \bmat{I_d}) $ is upper bounded by $(1 + \lambda)$, the NACF highlights the $(h,r)$ pairs at which the projected vectors point in the direction of the eigenvector associated to the largest eigenvalue of $(\mathbf{W} \odot \bmat{S}_r + \bmat{I_d})$, hence, the NACF is close to $\pm 1$ when the network weights and $\rstage$ adjacency are such that the magnitude of projections compared in (\ref{eq: nacf inequality}) are approximately the same. In that extreme case, future observations can be seen as eigenvectors of the network structure matrix $(\mathbf{W} \odot \bmat{S}_r + \bmat{I_d})$.

\section{Modelling COVID-19 ventilation beds admissions}
\label{sec: application}
Define the multivariate time series
$Y_{i, t}$ to be the recorded
number of COVID-19 patients transferred to mechanical ventilation beds
for day $t$, for NHS Trust $i$, across $d=140$ trusts in England. 
The $T=452$ \mbox{$d$-dimensional}
observations were obtained from the UK Government Coronavirus Dashboard
({\tt coronavirus.data.gov.uk}) between April 2020 and July 2021.
Patients who are transferred to mechanical ventilation beds
are a subset of those COVID-positive patients who are, or become, seriously ill and require
artificial ventilation. Some of these patients are directly transferred
in on ventilation from other hospitals and care settings.
\par
The $Y_{i, t}$ observations are count data and it is not appropriate to use regular GNAR
models directly and so, to stabilise variance and bring the data closer to normality, we study
the well-known and understood transformed version:
$X_{i, t} := \log ( 1 + Y_{i, t})$. We stack the individual trust $X_{i, t}$
time series and denote $\boldsymbol{X}_t = (X_{1, t}, \cdots, X_{140, t} ),$
for $t=1, \ldots, T=452$.
\par
Our network was built using  geographical `as the crow flies'
distances between NHS trusts with an optimal choice (i.e., best MSE predictive distance for one-stage neighbourhood regression) of 120km (approx 75~miles) between connecting nodes, hence, each trust is connected to trusts within a 120km radius of itself. The $X_{i, t}$ series for trust $i$ is associated
with vertex $i$ of our trust network depicted in Figure \ref{fig: rstage neighbors}a.
Figure~\ref{fig: correlation structure}a shows strong correlations between series at different nodes.
Our underlying assumption is that the nature of COVID-19 infection does not change drastically within England, and that the needs of similar hospitals can be described by the parsimonious GNAR model, which enables us to exploit the underlying structure.
Hence, we propose modelling $\boldsymbol{X}_t$ as a stationary {\em global-$\alpha$} GNAR process~\eqref{eq: vector wise}.

This GNAR parsinomity allows us to use significantly
more observations per parameter when estimating the autoregressive coefficients rather than separately for each node,
as would be required
for a (overparameterized) VAR model. 
For instance, fitting a VAR$(p)$ to $\boldsymbol{X}_t$ requires estimating (or, at the very least, having to consider) $p 140^2=19600p$ unknown parameters, whereas, fitting a $\GNAR(p, [s_1, \dots, s_p])$ requires estimating $p + \sum_{k= 1}^{p} s_k$ parameters, which is upper bounded by $p (1 + 6)$ given that the maximum $\rstage$ per lag is six and we are modelling the data as a global-$\alpha$ GNAR process.


\subsection{GNAR Model Selection}
\begin{figure}
        \centering
        \begin{subfigure}{0.48\textwidth}
            \includegraphics[width=\textwidth]{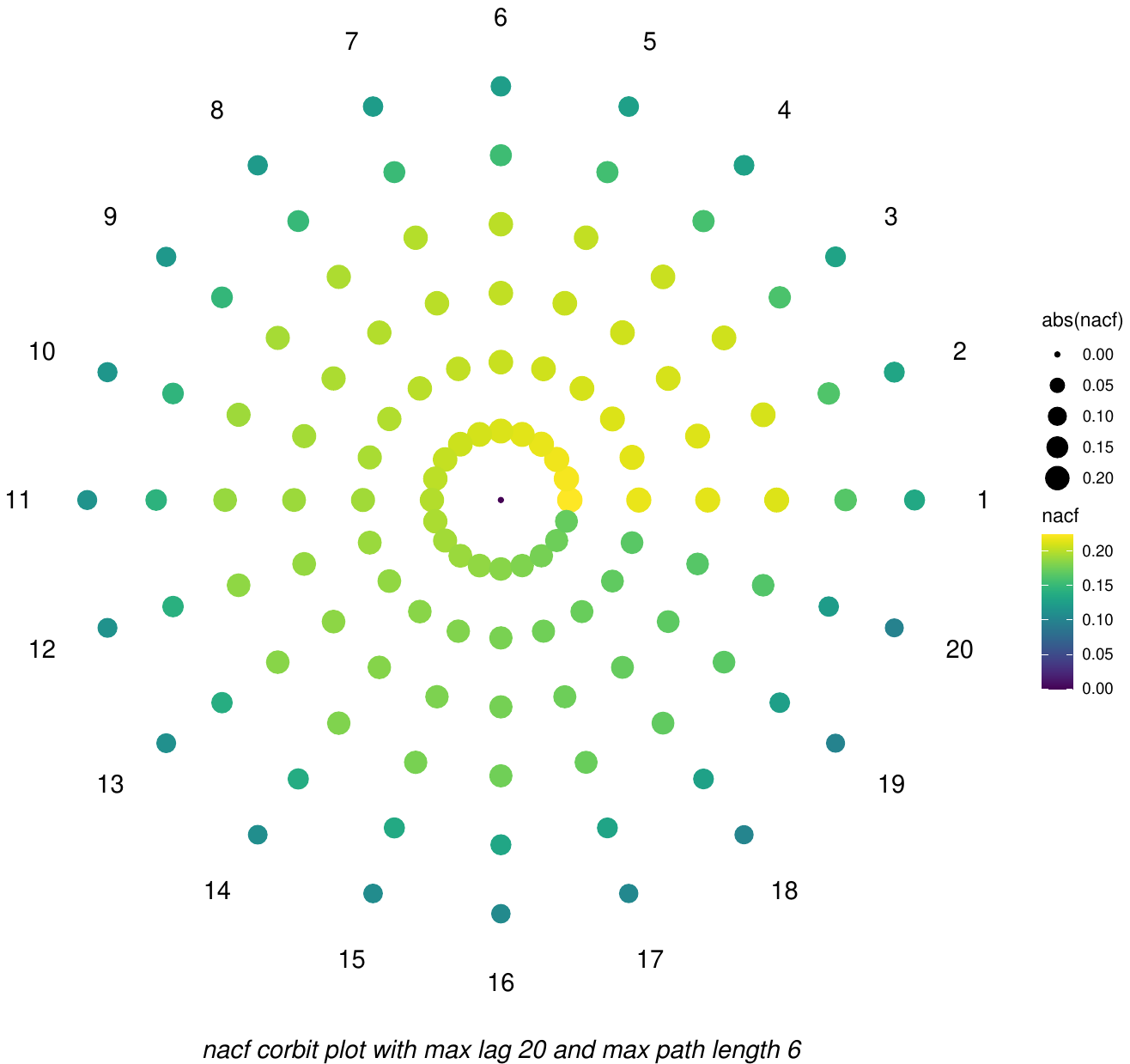}
            \caption{COVID-19 series NACF Corbit plot.}
            \label{fig: corbit nacf}
        \end{subfigure}
        \hfill
        \begin{subfigure}{0.48\textwidth}
            \includegraphics[width=\textwidth]{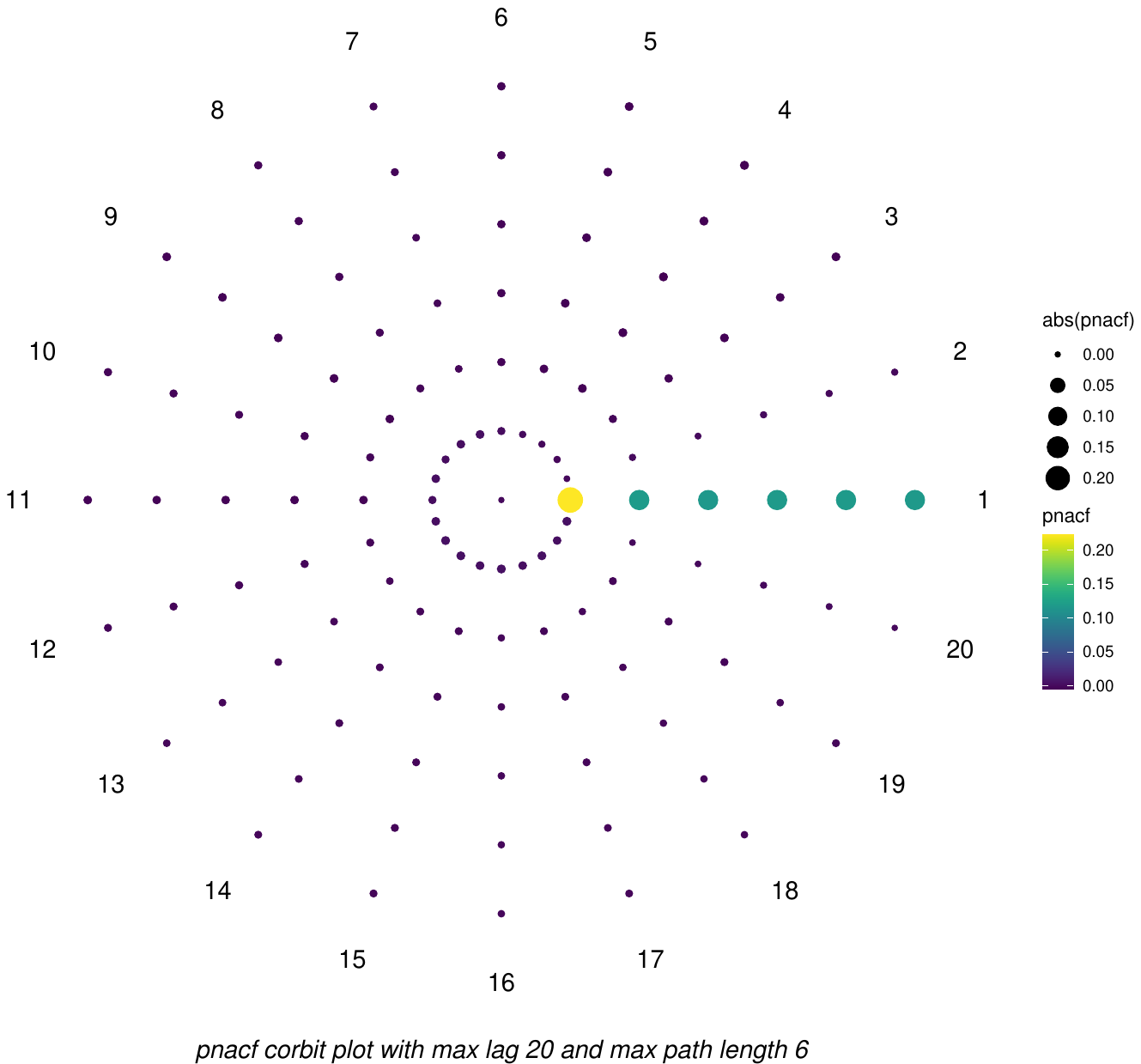}
            \caption{COVID-19 series PNACF Corbit plot.}
            \label{fig: corbit pnacf}
        \end{subfigure}
        \caption{Corbit plots for (a) observed NACF, and (b) for observed PNACF with respect to the COVID-19 series; maximum lag is equal to 20 and maximum $\rstage$ depth is equal to six. Plot (b) suggests a strong autoregressive behavior, which cuts-off after the first lag.}
        \label{fig: corbits}
    \end{figure}

Using the ideas from Section~\ref{sec:corbit}, we focus on the correlation structure in $\{ X_{i, t} \}$ and employ Corbit plots as graphical aids for assisting model selection. The observed NACF and PNACF values for our COVID-19 network time series are shown in Figures~\ref{fig: corbits}(a) and (b). 
Figure~\ref{fig: corbit pnacf} shows that the partial network autocorrelation cuts-off after the first time lag
and sharply decays after the first stage at the first time lag.
It remains roughly constant at the first lag for stages larger than or equal to two at lag~$1$, and cuts-off sharply at all stages for all lags $h \geq 2$.
Hence, our Corbit plot suggests looking at $1$-lag models for which there does not appear to be a large contribution to autocorrelation from $\rstage$ neighbors beyond the first stage. Looking at Corbit plots is considerably faster than a full evaluation of information criteria values such as
AIC or BIC, as mentioned earlier.
\par
We proceed by fitting a $\GNAR(1, [1])$ model and then compare this to both $\GNAR(1, [6])$ and  $\GNAR(6, [6, \ldots, 6])$ models,
the latter being the one with the largest possible number of active $\rstage$ regressions with the same number of lags.
We compare results of different GNAR models by the looking at the  one- and two-step prediction performance. For the two-step
predictive performance each model is fitted to the first $447$ observations of the COVID-19 network time series,
and two-step prediction is performed by using the one-step prediction as a pseudo-observation.
\begin{table}[H]
    \centering
    \begin{tabular}{crrr} 
    &&\multicolumn{2}{c}{Mean-squared Prediction Error(MSPE)}\\
    Model & No.\ Parameters & One-Step & Two-Step\\ 
    \hline
    GNAR(1, [1]) & 2 & 5.09 & 8.81 \\ 
    GNAR(1, [6]) & 7 & 5.06 & 8.8 \\ 
    GNAR(6, [6, \ldots , 6]) & 42&  5.24 & 9.88
    \end{tabular}
    \caption{One- and two-step predictive performance for different model order choices.}
    \label{tab: gnar comparison}
\end{table}
Table~\ref{tab: gnar comparison} shows that the order-6 model is the least-best choice based on one- and two-step predictive
performance. The predictive errors for the remaining first order models are very similar with the
$\GNAR(1, [6])$ best for one-step predictive error, very closely followed by the $\GNAR(1, [1])$ model, which is best for
the two-step predictive performance, but there is little to choose between here.

\subsection{Comparison to Other Models}
The underlying network structure aids not only in proposing a parsimonious GNAR model, it also permits estimation of the autoregressive coefficients more accurately and, we conjecture, reduces generalization error. In the interests of interpretability and illustration we select the  $\GNAR(1, [1])$ model and compare it to other popular multivariate time series models.
Table~\ref{tab: prediction comparison} compares
the predictive performance of the $\GNAR(1, [1])$ fit  to VAR$(1)$, sparse VAR$(1)$, restricted VAR$(1)$ and decoupled AR$(1)$ models for the COVID-19 (network) time series by performing one-step prediction ten times; see supplementary material for details.
\begin{table}[H]
    \centering
    \begin{tabular}{crr}
        Model & MSPE(sd) & Mean No. Parameters \\
        \hline
         GNAR(1, [1]) &  7.4(2.98) & 2 \\
         Sparse VAR(1) & 11.4(2.80) & 3089 \\
         Res. VAR(1) & 10.6(2.48) & 3773 \\
         VAR(1) & 12.3(3.18) & 19600 \\
         AR(1) & 87.4(5.15) & 140
    \end{tabular}
    \caption{Mean number of parameters and one-step predictive performance for the COVID-19 (network) time series for five different
        time series models over ten predictions. The mean-squared
        prediction error (MSPE) standard deviation is shown within parenthesis.}
    \label{tab: prediction comparison}
\end{table}
The models are fitted using the \texttt{GNAR} package \citep{gnar_package} for the $\GNAR(1, [1])$, \texttt{VARS} package \citep{varspack} for fitting the VAR$(1)$ and the restricted VAR$(1)$, \texttt{forecast} package \citep{fcst_pack} for the $140$
individual AR$(1)$ models, and \texttt{sparsevar} \citep{sparsevar} for the Sparse VAR$(1)$. We restricted each model to one lag
for a fair comparison based on the autocorrelation analysis the Corbit plot in Figure \ref{fig: corbit pnacf} provides.
\par
Table \ref{tab: prediction comparison} demonstrates the superlative performance of the GNAR process model for this data set. In particular,
its MSPE is about $35\%$ smaller than Sparse VAR$(1)$ and almost $30\%$ smaller than restricted VAR$(1)$.
Further simulations
show that for further time steps ahead the performance is even better. Hence, for the right kind of data, i.e.\ the GNAR model is (at least
approximately) valid, then GNAR forecasting can be extremely competitive.
\par
Moreover, model parsinomity eases interpretation of the results: our analysis provides evidence
that mechanical ventilation beds admissions `spreads' mostly through direct trust neighbours,
and also on the previous number of beds occupied at the trust. The parameter estimates for the $\GNAR(1, [1])$ were
$\hat{\alpha}_1 \approx 0.95$ and $\hat{\beta}_{1, 1} = 0.043$, and they were both statistically significant at the $0.1\%$ level.

For completeness we also tried fitting local-$\alpha$ models and
predictive power
improvement is negligible. The results are shown
in the
supplementary material Tables~\ref{tab: pred 443} thru~\ref{tab: pred 452},
where even with an extra 139 parameters the improvement in predictive
error is never more than 3\%, often nearer 1 or 2\%, and in two cases out
of the ten tables the simpler GNAR does better. This modelling
provides further validation for our initial global-$\alpha$ choice.

\subsection{Model Interpretation and Analysis}
\begin{figure}[H]
    \includegraphics[scale=0.35]{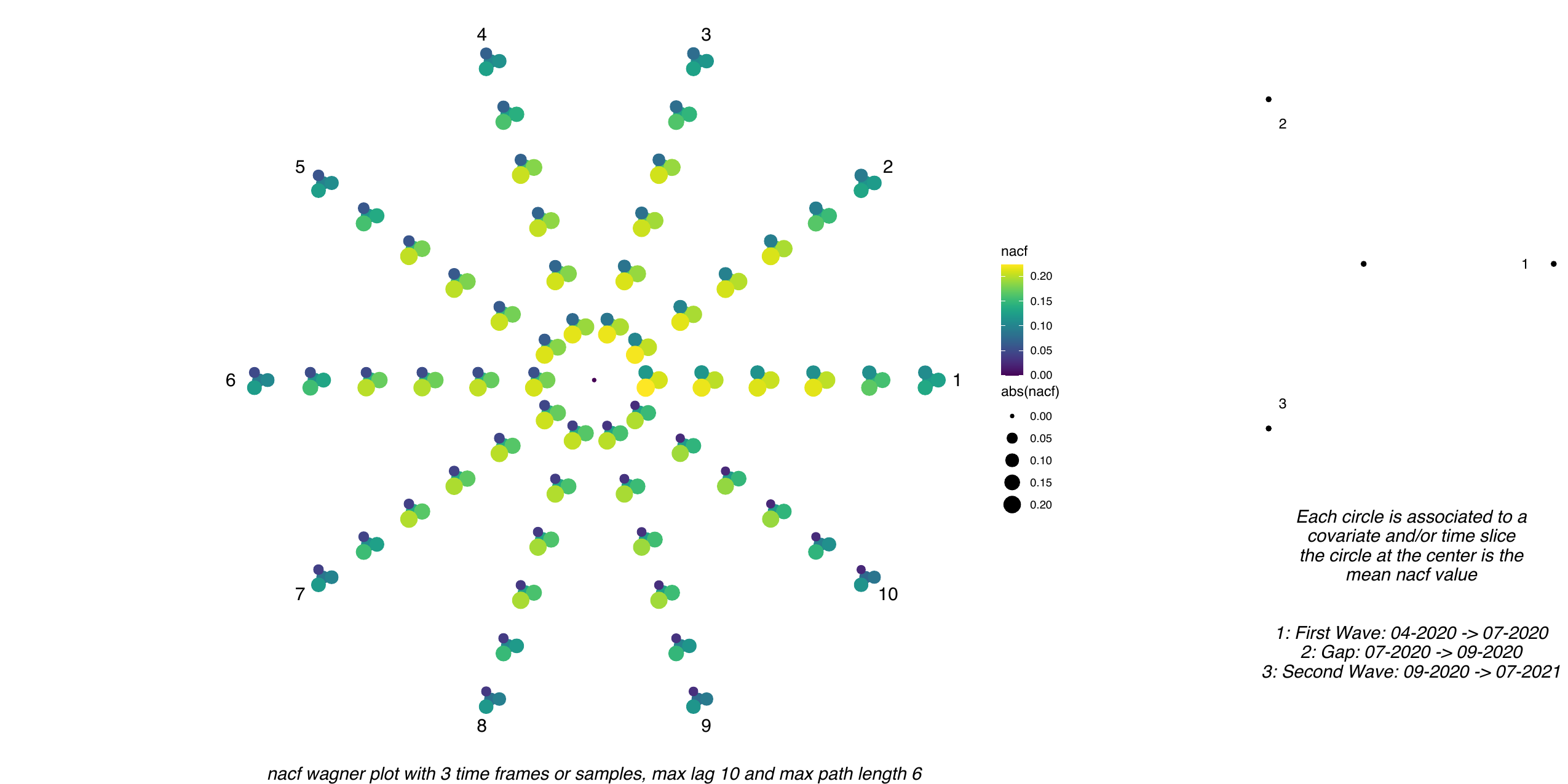}
        \caption{COVID-19 series NACF Wagner plot. The plot compares the NACF values between the two COVID-19 outbreaks and the gap between them.
        The maximum lag is ten and the maximum $\rstage$ depth is equal to six.  See text for description.}
    \label{fig: wagners}
\end{figure}

Another attractive property of GNAR models is that they can handle missing data and zero-values by weight readjustment; see \cite{gnar_paper}. We extend this idea to study the correlation structure of the two main COVID-19 outbreaks in England, the first outbreak was from April 2020 to July 2020 and the second one was from September 2020 to July 2021. We visualise the differences in the correlation structure during these different periods
by looking at Wagner plots, an example of which is shown in Figure \ref{fig: wagners}.
Wagner plots allow us to compare the NACF and PNACF values for different time-slices and/or covariates, we read the plot in the same manner as the Corbit plot and look at the legend on the right-hand side for distinguishing between covariates and/or time-slices. The point at the centre is the mean value of the NACF or PNACF values arising from the time-slices and/or covariate data splits. Essentially, if $c \in \{1, \dots, C\}$, where $C \in \mathbb{N}$ is the number of covariates or time-slices, then the value at the centre is
$\nacf(h, r) = C^{-1} \sum_{c = 1}^{C} \nacf_c (h, r),$ where $\nacf_c(h, r)$ is the NACF value corresponding to the covariate/time-slice $c$.  
\par
The Wagner plot in Figure \ref{fig: wagners} suggests that the correlation structure is different between the three time-slices, largest for the second outbreak and smallest for the gap. We interpret the gap period (July 2020 to September 2020) as an absorption state, which is characterised by constantly zero-valued observations, and perform model comparisons for the second outbreak in the table below. Prediction error is calculated using the last five realisations as test observations for the model fitted using the rest of the data.
\begin{table}[H]
    \centering
    \begin{tabular}{ccccc}
    & \multicolumn{2}{c}{Mean-squared Prediction Error} &&\\
     Model & One-Step & Two-Step & No. Parameters & Sig. Parameters \\ 
    \hline
    GNAR(1, [1]) & 5.43 & 12.74  & 2 & 2 \\ 
    GNAR(1, [6]) & 5.46 & 12.73 & 7 & 7 \\
    GNAR(6, [6, ..., 6]) & 5.87 & 12.99 & 42 & 20 
    \end{tabular}
    \caption{Comparison of different GNAR model orders for the series corresponding to the number of mechanical ventilation beds needed during the second COVID-19 wave for hospital trusts in the network shown by Figure \ref{fig: 1 stage neighbors}.}
    \label{tab: second outbreak comparison}
\end{table}
Table \ref{tab: second outbreak comparison} reflects the information the Corbit plots in Figure \ref{fig: corbits} provide by noting that models with more than one-lag should be discarded based on prediction error and the number of statistically significant parameters at the $1\%$ level. Moreover, it shows the slight performance improvements increasing $r^*$ from one to six has. Therefore, we, as previously, select GNAR(1, [1]), and exploit the ideas in Section \ref{sec: graph model connection} by introducing the following measures of relevance. 

\subsection{Node Influence}
\label{sec: node influence}
Once we fix an estimate of $r^*$, we can find the sparse weight matrix $\bmat{W} \odot \bmat{S}_{r^*}$ from which it is possible to compare the influence each node has. Motivated by (\ref{eq: nacf inequality}) we define the global relevance index as 
\begin{equation} \label{eq: global index}
    \globindex(X_{i, t}) := \bigg (\sum_{j = 1}^{d} [\bmat{W} \odot \bmat{S}_{r^*}]_{j i} \bigg )
   \bigg \{ \underset{l \in \mathcal{K}}{\max} \bigg ( \sum_{j = 1}^{d} [\bmat{W} \odot \bmat{S}_{r^*}]_{j l} \bigg ) \bigg \}^{-1},
\end{equation}
which computes the ratio between the largest column sum for active nodes and a particular node. We interpret this as the influence each node has globally on the correlation structure. 
\par
Next, we define a measure of the strength of pairwise interactions as the weighted contribution from all active $\beta_{kr}$ across active stages, formally,
\begin{equation} \label{eq: local influence}
    \local (i, j) := \bigg ( w_{ij} \sum_{k = 1}^{p} |\hat{\beta}_{kr}| \bigg ) \bigg \{ \sum_{l \in \mathcal{N} (i)} \sum_{r = 1}^{r^*} \sum_{k = 1}^{p}  w_{il} |\hat{\beta}_{kr}| \bigg) \bigg \}^{-1},
\end{equation}
where $\mathcal{N} (i) = \cup_{r = 1}^{r^*} \mathcal{N}_r (i)$, note that this results in $w_{ij}$ if we restrict the above to a specific $\rstage$. This index is larger as $w_{ij}$ gets closer to one.
Hence, the index computes the percentage of the neighbourhood coefficient across all lags corresponding to the specific pair (i.e., how strong, relative to the other active nodes, $j$ is for forecasting $i$).
\par
Also, by Theorem \ref{th: pop_sparsity gnar}, we can plot the correlation structure of our selected model based on the distances in the underlying network, which we plot by distinguishing between active nodes: $d(i, j) \leq r^*$, colliding nodes: $r^* < d(i,j) \leq 2 r^*$ and conditionally uncorrelated nodes: $d(i,j ) \geq 2r^* + 1$. Formally, we compute the relative strength of conditionally correlated nodes as
\begin{equation} \label{eq: conditional corr index}
    \rscc (i, j) := \{ d(i,j) \}^{-1} \mathbb{I} \{ d(i, j) \leq r^* \} + \{2 d(i,j) \}^{-1} \mathbb{I} \{ r^* < d(i, j) \leq 2 r^* \},
\end{equation}
where the distances are computed on the underlying network, and we divide by two the distance for nodes that are conditionally correlated despite not being active in their respective neighbourhood regressions.
These three measures of global and local influence, and a plot of the one-lag cross-correlation matrix are shown in Figure 6.
\begin{figure}
\begin{subfigure}{0.48\textwidth}
    \centering
    \includegraphics[width=\textwidth]{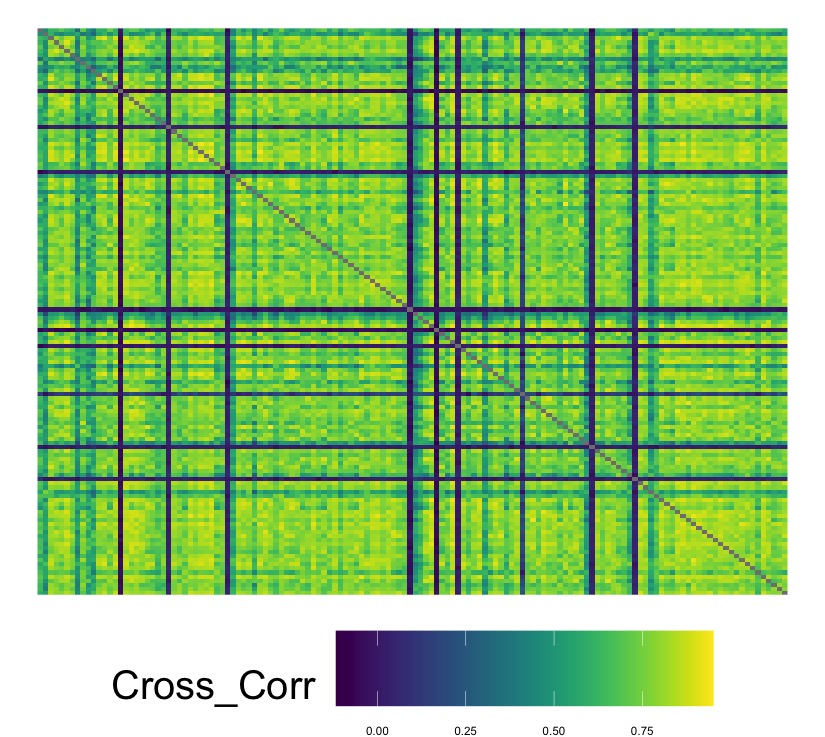}
    \caption{}
    \label{fig: cross-corr map}
\end{subfigure}
\hfill
\begin{subfigure}{0.48\textwidth}
    \centering
    \includegraphics[width=\textwidth]{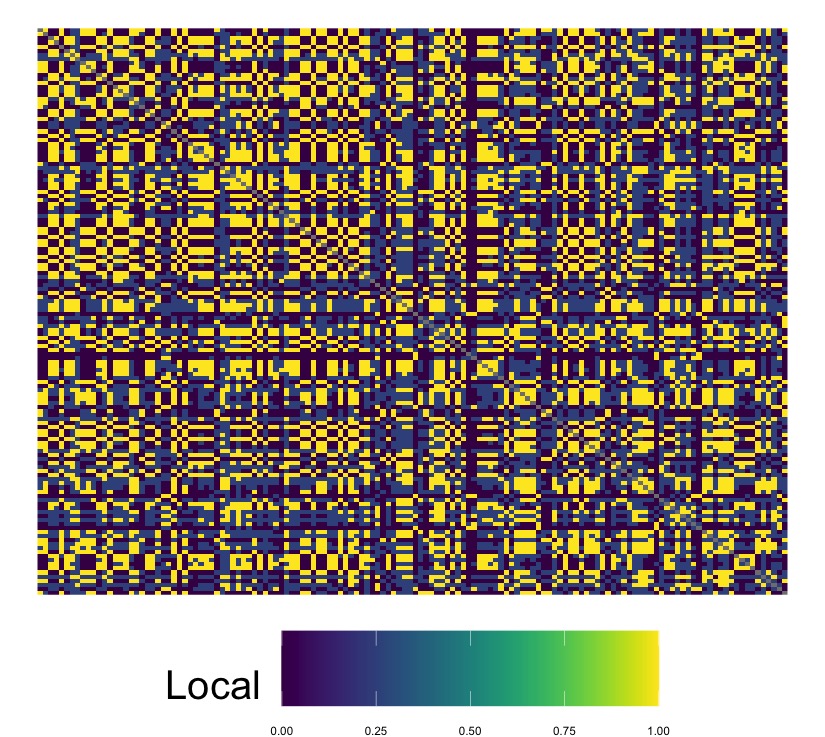}
    \caption{}
    \label{fig: local relevance}
\end{subfigure}
\hfill
\begin{subfigure}{0.48\textwidth}
    \centering
    \includegraphics[width=\textwidth]{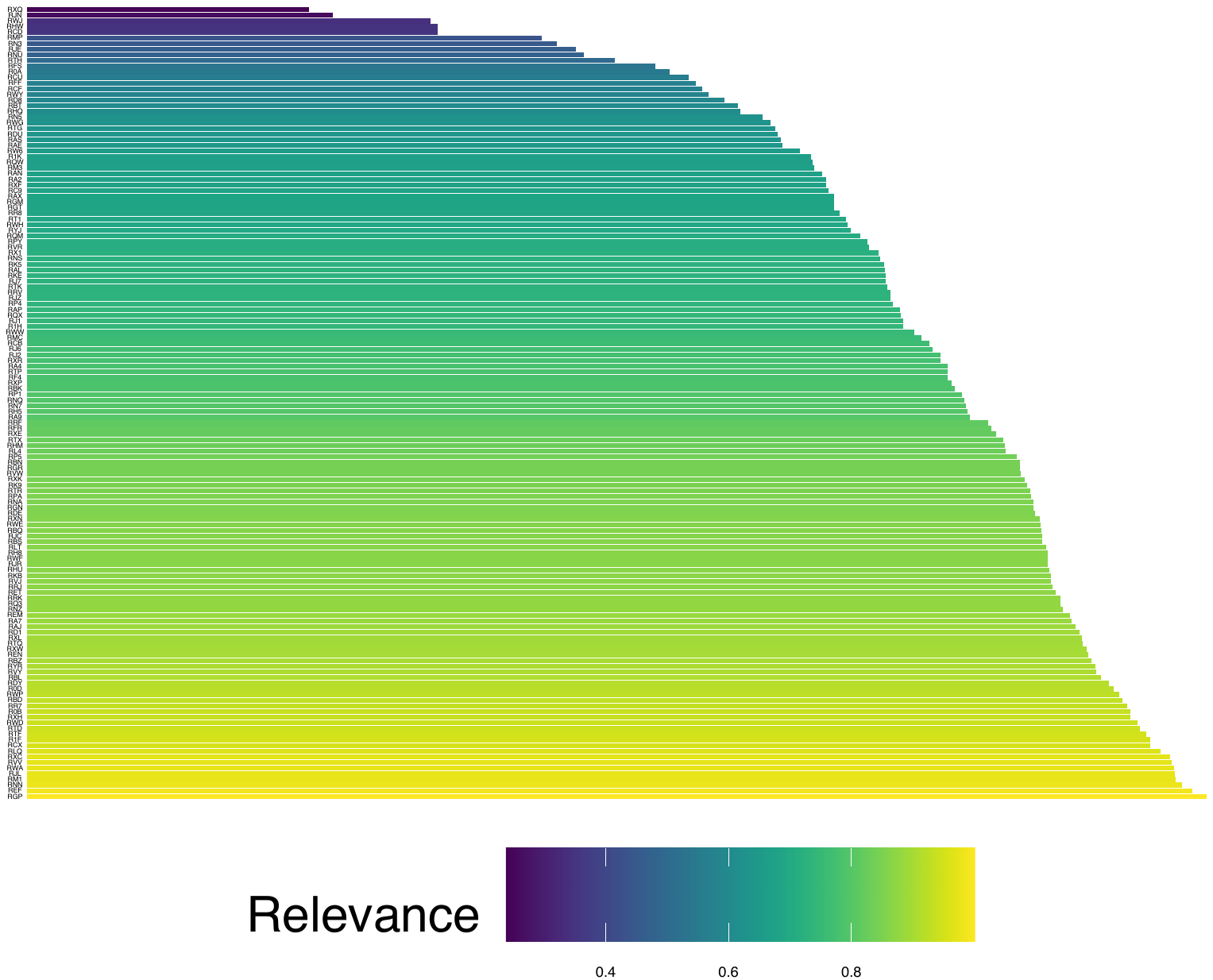}
    \caption{}
    \label{fig: global relevance}
\end{subfigure}
\hfill
\begin{subfigure}{0.48\textwidth}
    \centering
    \includegraphics[width=\textwidth]{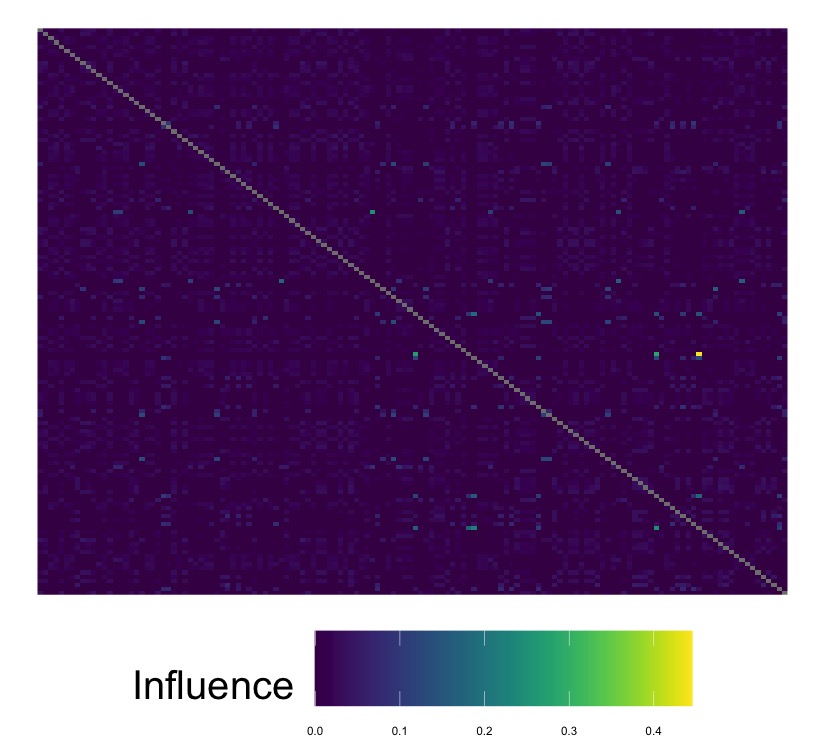}
    \caption{}
    \label{fig: active nodes}
\end{subfigure}
\caption{ \ref{fig: cross-corr map} shows a heat-map for the sample lag-one cross-correlations of the multivariate COVID series; \ref{fig: local relevance} highlights the structure of 1-stage neighbourhood regression across the entire network, where each strength of conditional correlation is given by $\rscc$ in~\eqref{eq: conditional corr index}; \ref{fig: global relevance} plots global node relevance defined by $\globindex$ in~\eqref{eq: global index}, which distinguishes nodes that have a large effect on the network correlation; \ref{fig: active nodes} shows  sparseness in active neighbourhood regressions by highlighting the strength of pairwise neighbourhood regression coefficients measured by $\local$ in~\eqref{eq: local influence}. The largest (yellow) coefficient on the right-hand side of
the plot corresponds to the transmission from the University Hospitals Plymouth to the Royal
Cornwall Hospitals NHS Trusts.}
\label{fig: correlation structure}
\end{figure}
Figure~\ref{fig: cross-corr map}  and \ref{fig: local relevance} compare the observed correlation with the
conditional correlation.
Both show clustering across the network in a self-similar manner, which is reflected by the strong autoregressive behaviour and dependence on one-stage neighbours.
Moreover, Figure \ref{fig: active nodes} shows the GNAR selection and shrinkage properties Section \ref{sub: subsect oracle selection} discusses by plotting a
(asymmetric) sparse heat-map of estimated neighbourhood regression coefficients, which highlights the relevance of key pair-wise interactions. 

\begin{figure}
\hspace{-2cm}
\begin{subfigure}{0.65\textwidth}
    \centering
    \includegraphics[width=\textwidth]{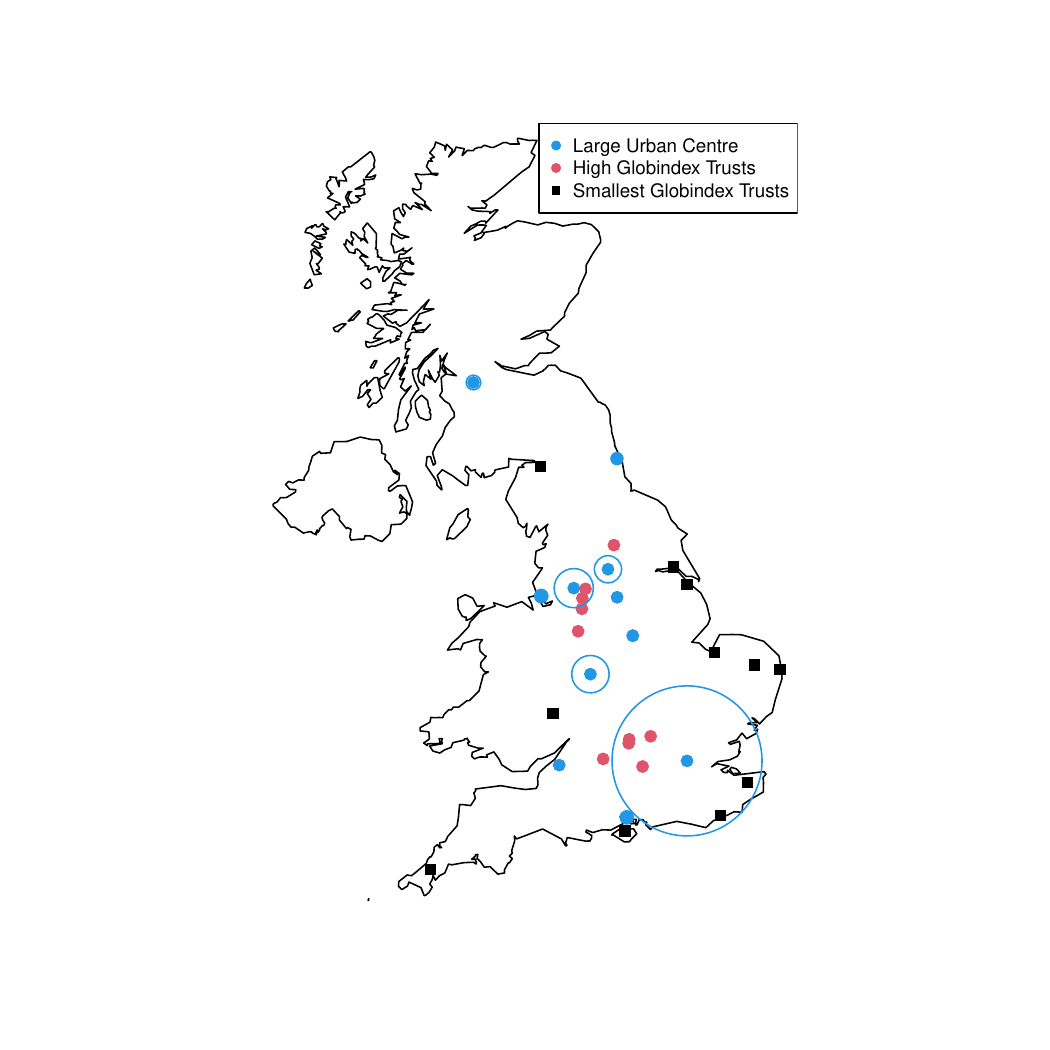}
    \caption{}
    \label{fig: conduitTowns10}
\end{subfigure}
\hspace{-2cm}
\begin{subfigure}{0.65\textwidth}
    \centering
    \includegraphics[width=\textwidth]{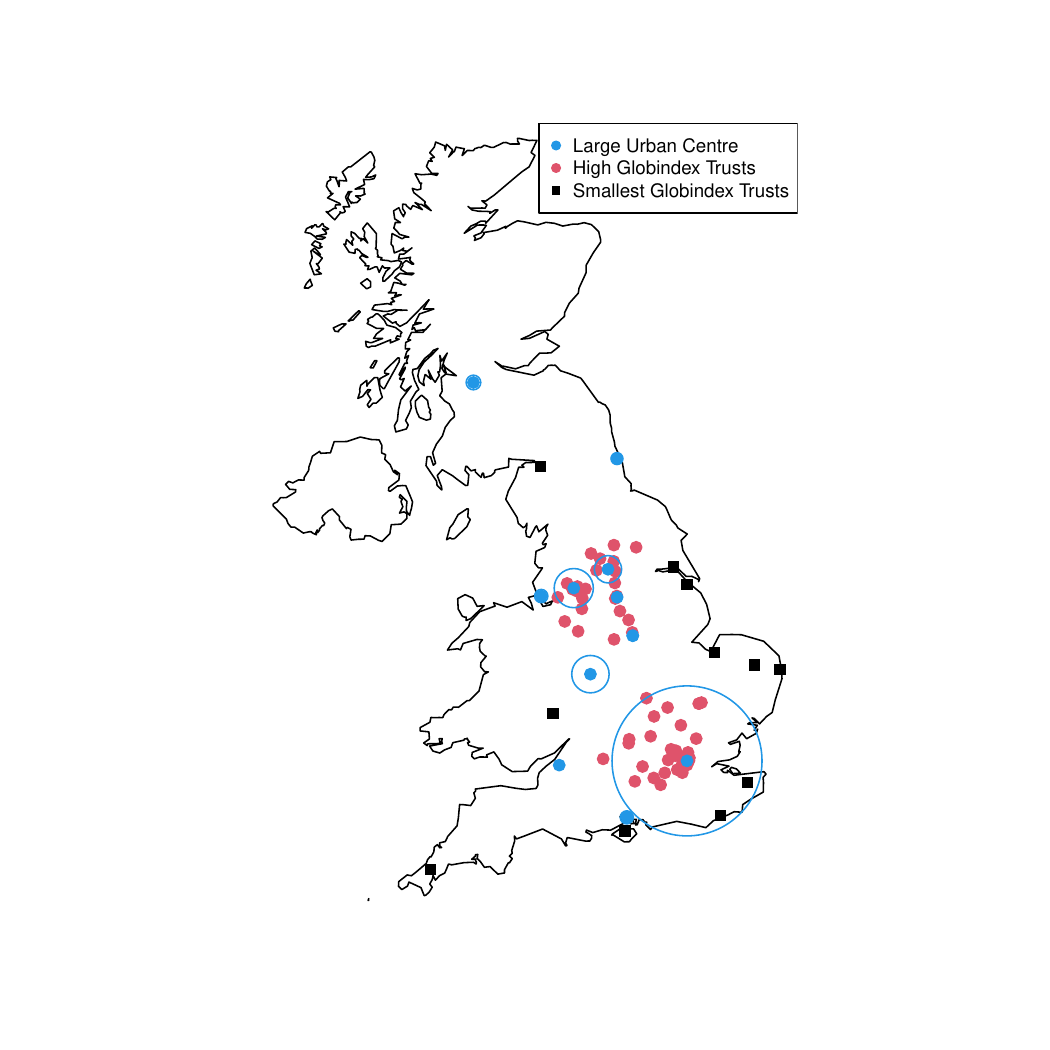}
    \caption{}
    \label{fig: conduitTowns60}
\end{subfigure}
\caption{Blue dots show 
top eleven urban centres in the UK.
Blue circles indicate urban centre population size with
London as approximately 9.78 million people.
Black squares are ten least `relevant' $\globindex$ Trusts.
Red dots: highest relevant NHS Trusts according to
$\globindex$. (a) top ten $\globindex$ Trust
locations (b) top sixty $\globindex$ Trust locations. \label{fig: conduitTowns}}
\end{figure}

Figure~\ref{fig: global relevance} shows the
$\globindex$ values for all NHS Trusts.
Individual Trust
identities are hard to discern from this
plot, but it is instructive
to examine them in a bit more detail.
The top ten
Trusts for $\globindex$ in decreasing order
are 1.\ Buckinghamshire,
2.\ East Cheshire,
3.\ Stockport,
4.\ Royal Berkshire,
5.\ Harrogate District, 
6.\ Tameside Glossop, 
7.\ Great Western (Swindon),
8.\ University Hospitals of North Midlands, 
9.\ Oxford Health and 
10.\ Oxford University Hospitals
with $\globindex$ influence ranging 
from $1.000$ to $0.952$.
Figure~\ref{fig: conduitTowns} shows the
most influential $\globindex$ Trusts for both the
first ten and first sixty Trusts in decreasing order.
It is interesting that the most influential Trusts
relating to the network time series
dynamics 
are located in two clear clusters: one positioned to the
north-west of London (between London, the West Midlands [Birmingham],
Bristol and the Southampton/Portsmouth urban centres)
and the other between and around the urban centres of the
West Midlands
[Birmingham], Manchester, Sheffield, Nottingham, West
Yorkshire and Liverpool. It is noticeable
that in the top ten $\globindex$ influence
list there are no Trusts in urban centres and only eighteen in the top sixty list. From the point of view
of network time series dynamics, the most influential trusts
are not urban centres {\em per se}, but intermediately-located Trusts. Having said this, it is intriguing
that there do not seem to be influential Trusts
immediately to
the south and east of Birmingham, nor between
Yorkshire and Tyneside, nor in the densely populated
area south and east of London.
The location of these influential Trusts might have
implications
for future epidemic mitigations
that could be taken to minimize
viral spread.

The least influential $\globindex$ Trusts
are those at the network extremities,
typically coastal towns and are shown as black squares
in Figure~\ref{fig: conduitTowns} (the exception
being the Wye Valley NHS Trust, which is an extremity
of the England NHS Trust network on the border with Wales).

\subsection{Discussion} \label{sub: application discussion}
An attractive property of GNAR parsinomity is that it enables us to interpret a large number of interaction with a relatively small number of parameters, in our case we can interpret the interactions between 140 NHS trusts with just two parameters
and {\em still} have superior forecasting performance, as shown
by Table~\ref{tab: gnar comparison}. The Corbit plots in Figure \ref{fig: corbits} show the strong autoregressive structure of the series $\boldsymbol{X}_t$, highlighting that correlation cuts-off after lag one. 
\par
 Table \ref{tab: prediction comparison} shows the excellent predictive performance of our chosen GNAR(1, [1]) model, which suggests that the outbreak is mostly dependent on the current node's past and gets worse by absorbing cases from one-stage neighbours, and, by Theorem \ref{th: pop_sparsity gnar}, nodes are conditionally uncorrelated from all other nodes given their two-stage neighbours --- i.e., communities can protect themselves by isolation.
\par
Further, and interestingly, the Wagner plot in Figure \ref{fig: wagners} shows that there is an increase in correlation during the second outbreak, which we attribute to less stringent measures and an increase in the number of interactions between nodes (i.e., more people travelling across different areas of England and Wales); see \cite{owidcoronavirus}. These results reflect the node relevance in Figure \ref{fig: global relevance}, which shows that the nodes with the most influence are close to London, and that the less influential nodes are at the corners, such as REF = Royal Cornwall Hospitals. Nevertheless, all nodes are similarly important given that we constrain the model to 1-stage neighbourhood regression, which shows the global properties of how demand for mechanical ventilation beds due to COVID-19 cases spread throughout England during the pandemic.

\section{Conclusion}
\label{sec:conc}
We have introduced a  new methodology for modelling and inferring relationships present in network time series data. Section \ref{sec:gnar methodology} reviews the GNAR model introduced by \cite{gnar_org, gnar_paper} and presents a new hierarchical representation, which can be written in compact matrix notation and is easier to manipulate. 
Subsequently, we proposed the NACF and the PNACF as diagnostic statistics, which can be visualised with Corbit plots. We note that these choices for measuring correlation are not the only possible ones, however, their usefulness and ease of interpretation as graphical aids for model selection are illustrated in Section \ref{sec: application}. Moreover, the ideas in Section \ref{sec: graph model connection} show the connection between GNAR models and graphical models, which suggest a clear interpretation of the relationship between nodes depending on their distance on the graph, and how this affects conditional correlation; see Theorem \ref{th: pop_sparsity gnar}. Also, we exhibit a possible interpretation of GNAR models as a constrained VAR that exploits prior knowledge of the network for performing variable selection and shrinkage, which enables us to study the connection between model constraints, prior information and parameter updates. These ideas allow us to study the local interactions between nodes, and the properties of the network as a composite object. Further, it exhibits how we can extend results from VAR theory to the GNAR framework by properly adjusting the correspondence between VAR parametrizations and GNAR formulations.
\par
We remark that the GNAR framework is useful for specific problems where the data can be properly described by an underlying network, which induces a particular correlation structure. Section \ref{sec: application} shows the advantages the GNAR methodology has when studying problems similar to the demand for mechanical ventilation beds during the COVID-19 pandemic. Also, motivated by Section \ref{sub: subsect oracle selection}, investigating the connection between prior information and posterior distributions, such as assigning a prior to the association function between nodes might improve forecasting performance and make complex models more interpretable.
\par
Future work will focus on trend removal, tests of stationarity and further developing the methodology we have presented. Also, a more thorough analysis of the NACF and PNACF might reveal interesting statistical properties for model selection, and aid in the study of AIC and BIC as criteria for parsimonious model selection; see \cite{aic, bic}. Moreover, based on the forecasting performance GNAR models have shown, it is worthwhile to study possible extensions by generalising GNAR models, for example, connecting the GNAR framework with generalised linear models.
\par
Essentially, GNAR is a parsimonious model that enables us to estimate fewer parameters more efficiently and precisely, even in high-dimensional settings. Furthermore, it is more transparent than overparametrized VAR, which facilitates interpretation and replication.
\par
The code for our plots and NACF/PNACF functions will be incorporated into the 
CRAN \texttt{GNAR} package in due course.

\section*{Acknowledgments}
We gratefully acknowledge the following support: Nason from EPSRC NeST Programme grant EP/X002195/1; 
Salnikov from the UCL Great Ormond Street Institute of Child Health, NeST, Imperial College London, 
the Great Ormond Street Hospital DRIVE Informatics Programme and the Bank of Mexico.
Cortina-Borja supported by the NIHR Great Ormond Street Hospital Biomedical Research Centre. The views expressed are those of the authors and not necessarily those of the EPSRC, National Health Service (NHS), the NIHR or the UK Department of Health.

\newpage

\begin{center}
{\large\bf SUPPLEMENTARY MATERIAL}
\end{center}

\begin{description}

\item[Correlation Structure Proof:]
Proof and illustration of Theorem \ref{th: pop_sparsity gnar}. (PDF/LaTeX)
\item[Network Autocorrelation Function Properties:]
Derivation and properties of the NACF given by Definition \ref{def: nacf}. (PDF/LaTeX)
\item[Algorithms and Properties of $\rstage$ Adjacency Matrices:]
Brief document exhibiting the properties and possible computation techniques for $\rstage$ adjacency matrices given by Definition \ref{def: rstage adjacency}. (PDF/LaTeX)
\item[R-package for  Graphical Aids routine:] R-package \textbf{GNAR} and scripts containing code to perform the diagnostic methods described in the article. The package also contains all data sets used as examples in the article. (GNU zipped tar file)

\item[COVID-19 Series data set:] Data set analyzed in Section 5. (.txt file)
\end{description}

\section*{Model Predictions}
We perform one-step prediction ten times to have a more robust estimate of out-of-sample prediction error, the summarised results are shown below.
\begin{table}[H]
    \centering
    \begin{tabular}{ccccc}
        Model & MSE(sd) & Mean No. Parameters \\
        \hline
         GNAR(1, [1]) &  7.414(2.977) & 2 \\
         Sparse VAR(1) & 11.307(2.768) & 2962 \\
         Res. VAR(1) & 10.614(2.482) & 3767 \\
         VAR(1) & 12.275(3.184) & 19600 \\
         AR(1) & 87.402(5.146) & 140 \\
         GNAR(1, [1])* & 7.313(2.900) & 141
    \end{tabular}
    \caption{One-step predictive performance and average number of parameters for the COVID-19 (network) time series for five different
        time series models. The MSE standard deviation is shown within parenthesis. GNAR(1, [1])* uses a different $\alpha_{id}$ for every NHS trust.}
    \label{tab: mode comp summary}
\end{table}
One-step prediction error (i.e., $||\hat{\boldsymbol{X}}_t - \boldsymbol{X}_t ||^2$) comparison between model predictions and the known test values. We perform the experiment ten times for $t = 443, \dots, 452$, and show the standard deviation within parenthesis next to each MSE value. The third column indicates the average number of active (i.e., non-zero coefficients) each model has throughout the experiment.
\par
Interestingly, GNAR parsinomity not only results in the smallest one-step prediction error in this case, it also has the smallest variance and standard deviation. Moreover, it is possible to fit more complex GNAR models since the number of unknown parameters is drastically smaller than for the other models, most notably, fitting a GNAR(1, [1]) requires estimating two parameters, whereas sparse VAR(1) requires estimating, on average, 3061 coefficients. 
\par
Remarkably, GNAR(1, [1]) accomplishes a smaller squared prediction error than sparse VAR(1) does with much fewer parameters.
\subsection*{Model Comparison for one-lag differenced series}
\begin{table}[H]
    \centering
    \begin{tabular}{ccccc}
        Model & MSE & Mean Active Parameters \\
        \hline
         GNAR(1, 1) &  7.847(2.908) & 2 \\
         Sparse VAR(1) & 7.686(2.940) & 451 \\
         Res. VAR(1) & 9.149(3.068) & 1840 \\
         VAR(1) & 11.371(3.525) & 19600 \\
         AR(1) & 7.816(2.931) & 140 \\
    \end{tabular}
    \caption{Mean number of parameters and one-step predictive performance for the differenced COVID-19 (network) time series for five different
        time series models. The MSE standard deviation is shown within parenthesis.}
    \label{tab: mode comp summary differenceed}
\end{table}
\subsection*{Prediction Error Tables for the Ten Experiments}
Below are the prediction error tables for each of the test points $\boldsymbol{X}_t$, where $t = 443, \dots, 452$, and  GNAR(1, [1])* uses a different $\alpha_{id}$ for every NHS trust. 
\begin{table}[H]
    \centering
    \begin{tabular}{ccc}
 Model & Active Parameters & One-Step SPE \\ 
\hline
GNAR(1, [1]) & 2 & 5.441 \\
 VAR(1) & $140^2$ & 11.613 \\
 Res. VAR(1) & 3821 & 9.902 \\
 Sparse VAR(1) & 2980 & 10.212 \\
 AR(1) & $140$ & 82.055 \\
 GNAR(1, [1])* & 141 & 5.324 
\end{tabular}
    \caption{One-step prediction error, $\hat{\boldsymbol{X}}_{443}$ is predicted using the previous 442 observations.}
    \label{tab: pred 443}
\end{table}

\begin{table}[H]
    \centering
    \begin{tabular}{ccc}
 Model & Active Parameters & One-Step SPE \\ 
\hline
GNAR(1, [1]) & 2 & 7.86 \\
 VAR(1) & $140^2$ & 12.804 \\
 Res. VAR(1) & 3712 & 11.504 \\
 Sparse VAR(1) & 2992 & 11.533 \\
 AR(1) & $140$ & 78.748 \\
 GNAR(1, [1])* & 141 & 7.833 
\end{tabular}
    \caption{One-step prediction error, $\hat{\boldsymbol{X}}_{444}$ is predicted using the previous 443 observations.}
    \label{tab: pred 444}
\end{table}

\begin{table}[H]
    \centering
    \begin{tabular}{ccc}
 Model & Active Parameters & One-Step SPE \\ 
\hline
GNAR(1, [1]) & 2 & 10.286 \\
 VAR(1) & $140^2$ & 12.52 \\
 Res. VAR(1) & 3730 & 11.605 \\
 Sparse VAR(1) & 2785 & 12.986 \\
 AR(1) & $140$ & 82.571 \\
 GNAR(1, [1])* & 141 & 10.173 
\end{tabular}
    \caption{One-step prediction error, $\hat{\boldsymbol{X}}_{445}$ is predicted using the previous 444 observations.}
    \label{tab: pred 445}
\end{table}

\begin{table}[H]
    \centering
    \begin{tabular}{ccc}
 Model & Active Parameters & One-Step SPE \\ 
\hline
GNAR(1, [1]) & 2 & 12.953 \\
 VAR(1) & $140^2$ & 17.762 \\
 Res. VAR(1) & 3812 & 14.472 \\
 Sparse VAR(1) & 3283 & 15.057 \\
 AR(1) & $140$ & 86.781 \\
 GNAR(1, [1])* & 141 & 12.676 
\end{tabular}
    \caption{One-step prediction error, $\hat{\boldsymbol{X}}_{446}$ is predicted using the previous 455 observations.}
    \label{tab: pred 446}
\end{table}

\begin{table}[H]
    \centering
    \begin{tabular}{ccc}
 Model & Active Parameters & One-Step SPE \\ 
\hline
GNAR(1, [1]) & 2 & 7.525 \\
 VAR(1) & $140^2$ & 13.82 \\
 Res. VAR(1) & 3761 & 10.815 \\
 Sparse VAR(1) & 3315 & 11.995 \\
 AR(1) & $140$ & 89.097 \\
 GNAR(1, [1])* & 141 & 7.431 
\end{tabular}
    \caption{One-step prediction error, $\hat{\boldsymbol{X}}_{447}$ is predicted using the previous 446 observations.}
    \label{tab: pred 447}
\end{table}

\begin{table}[H]
    \centering
    \begin{tabular}{ccc}
 Model & Active Parameters & One-Step SPE \\ 
\hline
GNAR(1, [1]) & 2 & 5.088 \\
 VAR(1) & $140^2$ & 12.504 \\
 Res. VAR(1) & 3830 & 9.976 \\
 Sparse VAR(1) & 3609 & 10.169 \\
 AR(1) & $140$ & 88.181 \\
 GNAR(1, [1])* & 141 & 4.989 
\end{tabular}
    \caption{One-step prediction error, $\hat{\boldsymbol{X}}_{448}$ is predicted using the previous 447 observations.}
    \label{tab: pred 448}
\end{table}

\begin{table}[H]
    \centering
    \begin{tabular}{ccc}
 Model & Active Parameters & One-Step SPE \\ 
\hline
GNAR(1, [1]) & 2 & 4.44 \\
 VAR(1) & $140^2$ & 7.978 \\
 Res. VAR(1) & 3738 & 7.716 \\
 Sparse VAR(1) & 3030 & 7.918 \\
 AR(1) & $140$ & 91.004 \\
 GNAR(1, [1])* & 141 & 4.466 
\end{tabular}
    \caption{One-step prediction error, $\hat{\boldsymbol{X}}_{449}$ is predicted using the previous 448 observations.}
    \label{tab: pred 449}
\end{table}

\begin{table}[H]
    \centering
    \begin{tabular}{ccc}
 Model & Active Parameters & One-Step SPE \\ 
\hline
GNAR(1, [1]) & 2 & 3.061 \\
 VAR(1) & $140^2$ & 7.196 \\
 Res. VAR(1) & 3759 & 5.888 \\
 Sparse VAR(1) & 3017 & 6.842 \\
 AR(1) & $140$ & 88.066 \\
 GNAR(1, [1])* & 141 & 3.089 
\end{tabular}
    \caption{One-step prediction error, $\hat{\boldsymbol{X}}_{450}$ is predicted using the previous 449 observations.}
    \label{tab: pred 450}
\end{table}

\begin{table}[H]
    \centering
    \begin{tabular}{ccc}
 Model & Active Parameters & One-Step SPE \\ 
\hline
GNAR(1, [1]) & 2 & 8.868 \\
 VAR(1) & $140^2$ & 10.869 \\
 Res. VAR(1) & 3772 & 10.999 \\
 Sparse VAR(1) & 3016 & 11.68 \\
 AR(1) & $140$ & 91.151 \\
 GNAR(1, [1])* & 141 & 8.607 
\end{tabular}
    \caption{One-step prediction error, $\hat{\boldsymbol{X}}_{451}$ is predicted using the previous 450 observations.}
    \label{tab: pred 451}
\end{table}

\begin{table}[H]
    \centering
    \begin{tabular}{ccc}
 Model & Active Parameters & One-Step SPE \\ 
\hline
GNAR(1, [1]) & 2 & 8.623 \\
 VAR(1) & $140^2$ & 15.693 \\
 Res. VAR(1) & 3794 & 13.258 \\
 Sparse VAR(1) & 3299 & 15.715 \\
 AR(1) & $140$ & 96.361 \\
 GNAR(1, [1])* & 141 & 8.546 
\end{tabular}
    \caption{One-step prediction error, $\hat{\boldsymbol{X}}_{452}$ is predicted using the previous 451 observations.}
    \label{tab: pred 452}
\end{table}

\bibliographystyle{guy3}
\bibliography{Bibliography-MM-MC}

\end{document}